\documentclass[prl,twocolumn,
amsfonts,superscriptaddress,showpacs,longbibliography]{revtex4-2} 
\usepackage{array}
\usepackage{amsmath,amssymb,mathrsfs}

\usepackage{natbib}
\usepackage{listings}
\usepackage{color}
\usepackage{multirow}
\usepackage[utf8]{inputenc}
\usepackage{comment}

\definecolor{dkgreen}{rgb}{0,0.6,0}
\definecolor{gray}{rgb}{0.5,0.5,0.5}
\definecolor{mauve}{rgb}{0.58,0,0.82}

\usepackage[normalem]{ulem}

\usepackage{cancel}
\usepackage{latexsym}
\usepackage{graphicx} 
\usepackage{epstopdf}
\usepackage{graphicx,epstopdf,color}
\usepackage{amsfonts}
\usepackage{hyperref}
\usepackage[dvipsnames]{xcolor}
\usepackage{url}
\usepackage{soul}
\usepackage{cancel}
\usepackage{appendix}
\usepackage{physics}
\usepackage{cleveref}

\usepackage{tikz}  
\usetikzlibrary{shapes.geometric}   
\usepackage{rotating}  
\usepackage{tikz-cd}
\usetikzlibrary{decorations.pathmorphing}
\usepackage{tikz-cd}
\usetikzlibrary{decorations.pathmorphing}
\usetikzlibrary{arrows,
    chains,
    decorations.markings,
    shadows, shapes.arrows,shapes, fit}

\tikzstyle{tensor}=[rectangle,draw=blue!50,fill=blue!20,thick]
\tikzset{%
dotted_block/.style={draw=black!20!white, line width=1pt, dash pattern=on 3pt off 1pt,
            inner ysep=3mm,inner xsep=2mm, rectangle, rounded corners}
}
\tikzstyle{lefttriangle}=[isosceles triangle,draw=blue!50,fill=blue!20,thick]
\tikzstyle{righttriangle}=[isosceles triangle,draw=blue!50,fill=blue!20,thick,rotate around={180:(0,0)}]
\tikzstyle{diamond} = [rectangle, draw=red!80, fill=red!40, thick, rotate = 45]
\tikzstyle{mpotens} = [circle, draw=red!50, fill=red!20]

\usepackage{xcolor}
\definecolor{kaki_blais}{RGB}{161, 163, 112}
\definecolor{blue_blais}{RGB}{60, 92, 129}
\definecolor{azzurro_blais}{RGB}{133, 187, 216}

\tikzset{
    wigglyarrow/.style={
        very thick,
        ->,
        blue_blais,
        >=Stealth,
        opacity = 0.8,
        decorate,
        line cap=round,
        decoration={
          snake,
          amplitude=1.mm,
          segment length=2.5mm,
          pre length=0mm,   
          post length=2.5mm}
    }
}

\crefname{equation}{Eq.}{Eqs.}
\Crefname{equation}{Equation}{Equations}
\crefname{figure}{Fig.}{Figs.}
\Crefname{figure}{Figure}{Figures}
\crefname{section}{Sec.}{Secs.}
\crefname{subsection}{Subsec.}{Subsecs.}
\Crefname{section}{Section}{Sections}
\crefname{appendix}{Appendix}{Apps.}
\Crefname{appendix}{Appendix}{Apps.}
\crefname{paragraph}{Sec.}{Secs.}
\crefname{table}{Table}{Tables}

\usepackage{bm}
\newcommand{\textalert}[1]{}

\newcommand{\eps}{\varepsilon}

\newcommand{\half}{\frac{1}{2}}

\usepackage{xr}

\graphicspath{{nb_fig/}}

\usepackage[caption=false]{subfig}
\usepackage{epstopdf}

\newcommand{\RN}[1]{
\textup{\uppercase\expandafter{\romannumeral#1}}
}

\makeatletter
\def\@fnsymbol#1{\ensuremath{\ifcase#1\or * \or \mathsection\or \mathparagraph\or \|\or **\or \else\@ctrerr\fi}}
\makeatother

\begin{document}
\title{First-principles study of dispersive readout in circuit QED}
\author{Angela Riva}
\thanks{ariva@lpthe.jussieu.fr}
\affiliation{Laboratoire de Physique de l’Ecole Normale Supérieure, Mines Paris, Inria, CNRS, ENS-PSL, Sorbonne Université, PSL Research University, Paris, France}
\author{Prakritish Gogoi} 
\affiliation{Sorbonne Université, CNRS, Institut des NanoSciences de Paris, 4 place Jussieu, 75005 Paris, France}
\author{Nicolas Gheeraert}
\affiliation{Krea University, Andhra Pradesh 517646, India}
\author{Serge Florens}
\affiliation{Univ. Grenoble Alpes, CNRS, Grenoble INP, Institut Néel, 38000 Grenoble, France}
\author{Alex W. Chin}
\affiliation{Sorbonne Université, CNRS, Institut des NanoSciences de Paris, 4 place Jussieu, 75005 Paris, France}
\author{Alain Sarlette}
\thanks{alain.sarlette@inria.fr}
\affiliation{Laboratoire de Physique de l’Ecole Normale Supérieure, Mines Paris, Inria, CNRS, ENS-PSL, Sorbonne Université, PSL Research University, Paris, France}
\author{Alexandru Petrescu}
\thanks{alexandru.petrescu@minesparis.psl.eu}
\affiliation{Laboratoire de Physique de l’Ecole Normale Supérieure, Mines Paris, Inria, CNRS, ENS-PSL, Sorbonne Université, PSL Research University, Paris, France}

\date{\today}
\begin{abstract}
The speed and fidelity of dispersive readout of superconducting qubits should improve by increasing the amplitude of the measurement drive. Experiments show, however, that beyond some drive amplitude there is always a saturation or drop in fidelity, often associated with a decrease in qubit energy relaxation time $T_1$. A simple Lindblad master equation does not capture the latter effect. More involved approaches based on effective master equations rely on strong assumptions about the spectra of the system and the bath and only partially agree with observations. Here, we perform a first-principles simulation of the full unitary dynamics of dispersive readout by considering the circuit QED Hamiltonian coupled to a microscopic model for the measurement transmission line, allowing for its arbitrary spectrum, including filters. Our access to the dynamics of the bath degrees of freedom allows us to investigate the emission spectrum of the system as a function of drive power. 
We show how the dependence of qubit $T_1$ on readout drive amplitude is sensitive to the details of the bath spectrum.
In particular, we find that $T_1$ drops with increasing drive amplitude when a Purcell notch filter is placed at the qubit frequency, and that the Lindblad master equation shows general qualitative defects compared to the first-principles model.
\end{abstract}
\maketitle

\paragraph{Introduction.---} 
Despite significant progress in speed and fidelity \cite{walter_et_al_2017,Sunada2022Apr}, qubit readout in superconducting circuit quantum electrodynamics (circuit QED) \cite{blais_et_al_2004,Blais_cQED} remains an important challenge on the path to fault-tolerant quantum computing, accounting for a significant part of the error budget in recent state-of-the-art implementations of the surface code \cite{Acharya2025Feb,Krinner2022May}. Implementing readout via analogs of cavity-QED Hamiltonians faces similar issues in other platforms, such as spin qubits  \cite{Muller2017Jan,Champain2025Mar,Burkard2023Jun}.

In the standard scheme for dispersive readout, a microwave resonator weakly and off-resonantly coupled to a superconducting qubit experiences a qubit-state-dependent frequency shift, which allows one to infer the state of the qubit by sending microwave signals (``drive'') close to the frequency of the resonator and measuring the output field. In this ideally quantum nondemolition (QND) measurement \cite{Braginsky1980Aug}, increasing the drive amplitude should render the measurement faster and higher fidelity. However, the observed measurement fidelity tends to drop beyond some drive amplitude, which signals the appearance of drive-induced non-QND behavior \cite{Sank_2016,walter_et_al_2017,Khezri2023Nov,Dai2025Jun}. 

\begin{figure}[t!]
    \centering

    \def\H{1.35}   
\usetikzlibrary{arrows.meta}

\tikzset{
    myarrow/.style={
        thick,
        <->,                
        >=Stealth,          
        shorten <=2pt,
        shorten >=2pt
    }
}
    \begin{tikzpicture}[scale=0.6]
    
        \foreach \x/\y in {-1.1/0.2} {
        \begin{scope}[shift={(\x,\y)}]
            \draw[very thick, kaki_blais, domain=-0.8:0.8, samples=100]  (-2.2, 2) -- (-1., 2);
            \draw[very thick, kaki_blais, domain=-0.8:0.8, samples=100]  (-2.2, 2.7) -- (-1., 2.7);
            \draw[line width=3pt, kaki_blais, domain=-0.8:0.8, samples=100] (-1.6, 2.35) circle (0.9cm);
    
            \fill[kaki_blais!40] (-1.6, 2.7) circle (0.2cm);
            \node[kaki_blais, ultra thick] at (-1.6, 1.) {$\omega_{q}$};  
            \node[ultra thick] at (-0.3, 3.9) {$g$};  
    
        \end{scope}
        }
    
      \draw[line width=3pt, blue_blais, domain=-0.8:0.8, samples=100] plot (\x, {1.6+3*\x*\x});
          \node[blue_blais, ultra thick] at (0, 3.7) {$\vdots$};  
          \node[blue_blais, ultra thick] at (0, 1.2) {$\omega_a$};  
          
      \foreach \y in {1.8, 2.1, 2.4, 2.7, 3} {
        \pgfmathsetmacro{\xlim}{sqrt((\y-1.6)/3)}
        \draw[blue_blais, opacity=0.5, very thick] (-\xlim,\y) -- (\xlim,\y);
      }
      \fill[blue_blais!40] (0, 2.1) circle (0.15cm);
      \fill[blue_blais!60] (0, 2.4) circle (0.15cm);
      \fill[blue_blais!20] (0, 2.7) circle (0.15cm);
      
      \draw[wigglyarrow] (0,-0.4) -- (0,1);
      \node[blue_blais] at (0.55, -0.2) {$\omega_d$};
    
      \foreach \x/\y in {2.5/0.1, 3.8/1.4, 3.4/3.05, 2.3/4.6} {
        \begin{scope}[shift={(\x,\y)}]
          \pgfmathsetmacro{\a}{0.5*\y + 2}
          \pgfmathsetmacro{\xmax}{sqrt(\H/\a)}
          
          \draw[azzurro_blais, domain=-\xmax:\xmax, samples=100, line width=2.3pt, opacity=1] plot (\x, {(0.5*\y+2)*\x*\x});
          \node[scale=0.8, azzurro_blais, ultra thick] at (0, 1.59) {$\vdots$};
          
          \foreach \mult in {3, 7, 11} {
              \pgfmathsetmacro{\yconst}{\mult * sqrt(\y+0.1)*0.05}
              \pgfmathsetmacro{\xlimr}{sqrt(\yconst/(0.5*\y + 2))}
              \draw[azzurro_blais, opacity=0.5, thick] 
                (-\xlimr,\yconst) -- (\xlimr,\yconst);
            }
        \end{scope}
        }
        
        \foreach \x/\y in {2.5/0.1,  3.8/1.4} {
        \begin{scope}[shift={(\x,\y)}]
        \foreach \mult in {15, 19} {
              \pgfmathsetmacro{\yconst}{\mult * sqrt(\y+0.1)*0.05}
              \pgfmathsetmacro{\xlimr}{sqrt(\yconst/(0.5*\y + 2))}
              \draw[azzurro_blais, opacity=0.5, thick] 
                (-\xlimr,\yconst) -- (\xlimr,\yconst);
            }
        \end{scope}
        }
    
        \foreach \x/\y in {2.5/0.1} {
        \begin{scope}[shift={(\x,\y)}]
        \foreach \mult in {23, 27, 31, 35, 39, 43} {
              \pgfmathsetmacro{\yconst}{\mult * sqrt(\y+0.1)*0.05}
              \pgfmathsetmacro{\xlimr}{sqrt(\yconst/(0.5*\y + 2))}
              \draw[azzurro_blais, opacity=0.5, thick] 
                (-\xlimr,\yconst) -- (\xlimr,\yconst);
            }
        \end{scope}
        }
    
        \coordinate (Q) at (-2.7, 3.08);
        \coordinate (A) at (0, 3.2);
        \coordinate (B1) at (2.8, 1.3);
        \coordinate (B2) at (3.5, 2.6);
        \coordinate (B3) at (3.4, 3.9);
        \coordinate (B4) at (2.3, 5.3);
        \coordinate (B5) at (0.6, 5.5);
    
        \node[scale=0.8, azzurro_blais, ultra thick] at (3.2, 0.2) {$\omega$};
        \node[scale=0.8, azzurro_blais, ultra thick] at (4.5, 1.6) {};
        \node[scale=0.8, azzurro_blais, ultra thick] at (4.1, 3.3) {};
        \node[scale=0.8, azzurro_blais, ultra thick] at (3., 5.) {};

    
        \draw[myarrow, bend left=50, opacity=0.7] (Q) to (A);
        \draw[myarrow, bend left=60, opacity=0.1] (A) to node[scale=0.8, pos=0.7, left, opacity=0.5] {$J(\omega)$} (B1);
        \draw[myarrow, bend left=40, opacity=0.2] (A) to node[scale=0.8, pos=0.7, below] {} (B2);
        \draw[myarrow, bend left=20, opacity=0.3] (A) to node[scale=0.8, pos=0.8, below] {} (B3);
        \draw[myarrow, bend left=-10, opacity=0.4] (A) to node[scale=0.8, pos=0.8, below] {} (B4);
        \draw[myarrow, bend left=-40, opacity=0.5] (A) to (B5);
    
        \node[azzurro_blais, ultra thick] at (0.6, 5.5) {$\dots$};
        
    \end{tikzpicture}
    
    \caption{ \emph{Microscopic model}: A qubit of transition frequency $\omega_{q}$ is Rabi-coupled, with strength $g$, to a readout resonator mode of frequency $\omega_a$, driven at frequency $\omega_d \simeq \omega_a$. The resonator is coupled to a bath of harmonic oscillators characterized by the bath spectral density function $J(\omega)$.}
    \label{fig:sketch}
\end{figure}
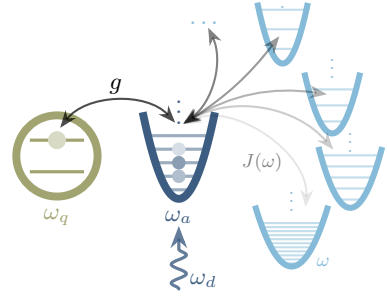

One symptom of non-QND behavior during measurement is the change in the measured qubit energy relaxation time $T_1$ as the drive amplitude is increased \cite{Slichter2012Oct, Mundhada2016,Gusenkova2020Sep,Hutin2024Oct,Bista2025Jan,May2025Feb}. In the ideal situation of dispersive readout \cite{Blais_cQED}, measuring the qubit state should only induce backaction in the form of dephasing. In the Lindblad master equation formalism, it has been shown that dephasing noise intrinsic to the qubit can be converted by the readout drive into additional relaxation for the qubit, resulting in a drop in $T_1$ with readout drive amplitude \cite{boissonneault_et_al_2009,Slichter2012Oct}; on the other hand, the $T_1$ induced by the Purcell effect has been shown to increase  \cite{boissonneault_et_al_2009,sete_gambetta} in models based on a two-level truncation of the qubit-encoding system, but can decrease in models based on multi-level weakly anharmonic systems \cite{alex_perturbation,ash_clerk_intrinsic}. These rates can be significantly changed by resonances occurring in the driven spectrum of the qubit system which also lead to leakage outside of the computational manifold, the so-called measurement-induced state transitions \cite{lescanne_et_al_2019,verney_et_al_2019,shillito_et_al_2022,Cohen2023Apr,Dumas2024Oct,Dai2025Jun}. The relaxation rates can also change as the qubit peak is ac Stark shifted and broadened by the readout drive, which brings it in resonance with additional spurious degrees of freedom in its bath \cite{thorbeck,Dai2025Jun}.

This points to the necessity of a model of the system-bath coupling which captures the frequency dependence of the bath. One possibility is to attempt to derive a Lindblad master equation from a microscopic model of the bath \cite{breuer_petruccione}. This approach relies on several approximations (see below) and is difficult in practice in dispersive readout: in order to derive the collapse operators, one would need to find the Floquet modes \cite{Blumel_87, Grifoni_Hänggi_1998} of the periodic system Hamiltonian comprising qubit, resonator, and readout drive. However, as the system is nearly resonantly driven, the problem becomes intractable \cite{Breuer_Holthaus_1989}. One 
therefore typically resorts
to phenomenological master equations, such as those based on a simple single-photon dissipator describing resonator relaxation \cite{blais_et_al_2004,Blais_cQED,shillito_et_al_2022}, where the frequency dependence of the spectral function is neglected.

In this Letter, we overcome these difficulties by numerically simulating the full unitary dynamics of a first-principles model of the circuit QED Hamiltonian coupled to an electromagnetic environment, as specified by the spectral density function \cite{Caldeira1983Sep}. As such, we do not make any of the standard approximations in the derivation of master equations \cite{breuer_petruccione}. Our tensor network approach \cite{Haegeman_2011} is nonperturbative in the frequency-dependent system-to-bath coupling, and does not use the secular or Markov approximations. Since we simulate the time evolution of the full wavefunction, we have access to the dynamics of the bath observables \cite{chin_chain_mapping}. By tracking which bath modes are being populated, we are able to extract the emission spectrum of the system. To isolate the effects due to the bath degrees of freedom, as opposed to drive-induced state transitions \cite{Dai2025Jun} inside of a multi-level system representing the qubit, we consider a \textit{two-}level system coupled to a driven harmonic readout resonator \cite{blais_et_al_2004,Blais_cQED}, and we extract the drive power dependence of the relaxation rate of the qubit, for arbitrary spectral distributions of the bath. This captures behaviors that are beyond the reach of previous models in the presence of complex filtered baths \cite{metamaterial_2017, metamaterial_2018, Putterman_2025, amazon_2025, walter_et_al_2017, Hutin2024Oct, nakamura_filter_intrinsic, xiao2025flexiblereadoutunconditionalreset, Reed_2010, Yan_2023}.\\

\begin{table}[t!]
  \begin{ruledtabular}
    \begin{tabular}{cc}
      \rule{0pt}{3ex} \textbf{$J(\omega)$} & {Parameter} \\ [1ex] \hline
      
      \rule{0pt}{6ex} 
      $J_{\mathrm{flat}}(\omega) = 2\alpha_{\rm flat} \Theta(\omega_{\rm max}-\omega) \Theta(\omega-\omega_{\rm min})$ & 
      $\begin{matrix} \omega_{\rm min} = 3 \\ \omega_{\rm max} = 12 \end{matrix}$ \\ [2ex]
      
      $J_{\mathrm{Ohm}}(\omega) = 2\alpha_{\rm Ohm}\omega\Theta(\omega_c-\omega)$ & 
      $\omega_c = 15$ \\ [2ex]
      
      $J_{\mathrm{PF}}(\omega) = J_{\mathrm{Ohm}}(\omega) \left(1 - D e^{-\frac{(\omega-\omega_q)^2}{2\sigma^2}}\right)$ & 
      $\begin{matrix} D = 0.7 \\ \sigma = 0.1 \end{matrix}$ \\
      \rule{0pt}{1ex}
    \end{tabular}
  \end{ruledtabular}
  \caption{Bath spectral functions considered in \cref{fig:t1_vs_nbar}. Parameter values are in units of $2\pi$\,GHz and $\Theta$ is the Heaviside step function. The value of the prefactor $\alpha$ for each case is selected to impose the corresponding single-photon decay rate of the resonator $\kappa = 2\pi J(\omega_a) = 50\text{ MHz}$.} 
  \label{tab:sdf_table}
\end{table}

\paragraph{Model.---} Our starting point is a Caldeira-Leggett type model \cite{Caldeira1983Sep}, which describes the bath as a collection of non-interacting harmonic oscillators coupled to the system ($\hbar=1$)
\begin{align}
\begin{split}
    \hat H(t) =& \hat H_S(t) \\&+ \int d\omega [ \sqrt{J(\omega)} (\hat a + \hat a^\dagger) (\hat b_\omega + \hat b_\omega^\dagger) +  \omega \hat{b}_\omega^\dagger \hat{b}_\omega ]. \label{eq:model_star}
\end{split}
\end{align}
In this model $\hat H_S(t) = \frac{\omega_{q}}{2} \hat \sigma_z + \omega_a \hat a^\dagger \hat a + g \hat \sigma_x(\hat a+\hat a^\dagger) + \lambda (\hat a+\hat a^\dagger)^2 + \eps_d \sin(\omega_d t) \left(\hat a + \hat a^\dagger \right)$ is a minimal circuit QED Hamiltonian describing a qubit ($\hat\sigma_z = \ketbra{1}-\ketbra{0}$, $\hat\sigma_x = \ketbra{1}{0}+\ketbra{0}{1}$) of transition frequency $\omega_q$ transversely coupled to a readout resonator modeled by a harmonic oscillator of frequency $\omega_a$, with coupling strength $g$. As needed for dispersive readout \cite{Blais_cQED}, the readout resonator is driven at a nearly resonant frequency $\omega_d\simeq \omega_a$, with drive amplitude $\eps_d$. We describe our procedure to determine this frequency below. The system Hamiltonian further contains a correction (term in $\lambda$) needed to enforce the gauge-invariant form of the system-to-bath coupling [SM]. 

In \cref{eq:model_star} the bath operators obey the canonical commutator $[ \hat b_\omega, \hat b_{\omega'}] = \delta(\omega -\omega')$, whereas the spectral function $J(\omega)$ encodes the strength of the coupling to bath modes at frequency $\omega$. An idealized form modeling a transmission line in circuit QED experiments would be Ohmic $J(\omega) \propto \omega$ \cite{Cattaneo_2021}, but in realistic experimental situations the electromagnetic environment of the circuit is much more complex \cite{Dai2025Jun,thorbeck, walter_et_al_2017, Hutin2024Oct, amazon_2025,Putterman_2025, Sunada2022Apr,xiao2025flexiblereadoutunconditionalreset}. In this work, exploiting the fact that our approach can handle an arbitrary spectral function $J(\omega)$, we will consider several forms to illustrate qualitatively different behaviors of qubit relaxation during readout, see Table~\ref{tab:sdf_table}.

To bring the model \cref{eq:model_star} to a form suitable for numerical simulation, we unitarily map the bath Hamiltonian to that of a bosonic tight-binding chain with nearest-neighbor hopping terms \cite{chin_chain_mapping}, that remains linearly coupled to the system 
\begin{align}
\begin{split}
\hat H(t)
=& \hat H_S(t) + k_0\,(\hat a+\hat a^\dagger)\,(\hat c_0+\hat c_0^\dagger )  \\ 
&+ \sum_{i=0}^\infty \Bigl(e_i\,\hat c_i^\dagger\hat c_i
   + t_i\,\hat c_{i+1}^\dagger\hat c_i + t_i\,\hat c_i^\dagger\hat c_{i+1}\Bigr).
   \label{eq:model_chain_2} 
\end{split}
\end{align}
On-site bosonic creation and annihilation operators obey the canonical commutator $[\hat{c}_i,\hat{c}_j^\dagger] = \delta_{ij}$ for all $i,j \geq 0$,  on-site energies are given by $e_i$, and hopping energies by $t_i$. 
The readout resonator is transversely coupled with an energy scale $k_0$ to the first site of the bath. Upon suitable truncation of the chain length, \cref{eq:model_chain_2} has an efficient matrix product operator representation \cite{SCHOLLWOCK201196} [SM].

\begin{figure}[t!]
    \centering    \includegraphics[width=0.95\linewidth]{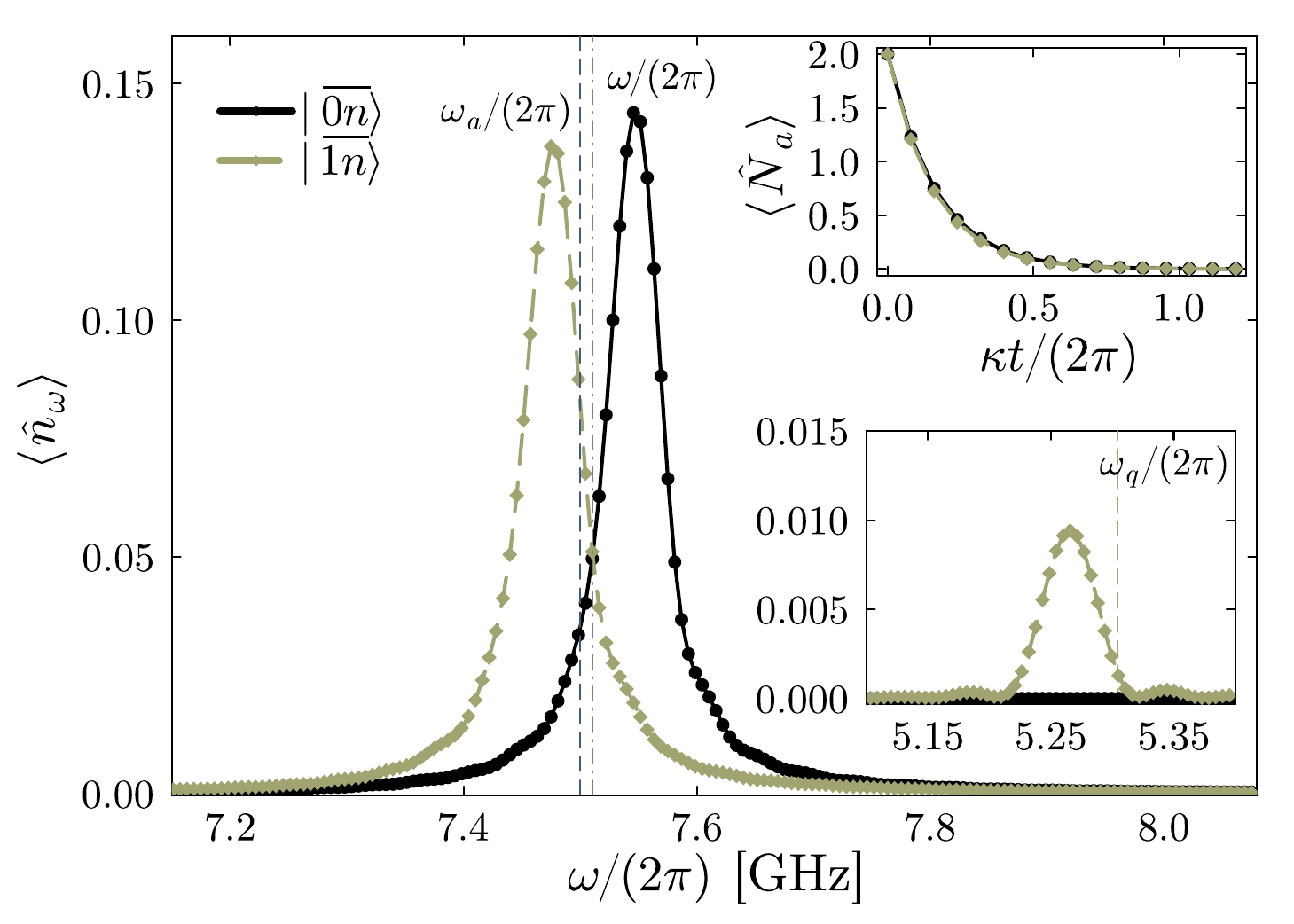}
    \caption{\textit{Numerical calibration (MPS simulation):} The main plot shows frequency resolved occupancies $\langle \hat n_\omega \rangle$ of the bath normal modes at time $\kappa t/(2\pi)=1$, sufficient to exponentially relax the resonator population $\langle \hat{N}_a \rangle$ (top right inset), for an Ohmic spectral density function $J_{\rm Ohm}(\omega)$. For initial conditions $\ket{\overline{1n}}$ and $\ket{\overline{0n}}$ with $n=2$, the respective peak positions $\omega_{a1} / (2\pi) =\arg\max_{\omega}\langle \hat n_\omega\rangle = 7.475 \, {\rm GHz}$ and $\omega_{a0} / (2\pi) = 7.546 \,{\rm GHz}$ correspond to the qubit-state-dependent resonator frequencies. The midpoint frequency $\bar\omega / (2\pi) =\big(\omega_{a1}+\omega_{a0}\big)/(4 \pi) = 7.51 \, {\rm GHz}$, shown as a vertical dot-dashed line, sets the readout drive frequency for our next simulations and differs from the bare resonator frequency $\omega_a / (2\pi)= 7.5  \, {\rm GHz}$ (dashed vertical line) due to Lamb shift. The bottom-right inset (same axes as main plot) shows a peak in $\expval{\hat n_\omega}$ around $\omega_q/(2\pi)$, corresponding to the emission of the qubit when initializing the system in $\ket{\overline{1n}}$.  
     \label{fig:calibration}
     }
\end{figure}

In our simulations, we initialize the system and bath in product states of the form
\begin{align}\label{eq:init}
    \ket{\psi(t=0)}=\ket{\overline{jn}} \ket{\mathbf{0}},
\end{align}
where $\ket{\mathbf{0}}$ is the vacuum of the uncoupled chain, i.e. $\hat{c}_i \ket{\mathbf{0}} = 0$, for $i=0,1,\ldots$ and $\ket{\overline{jn}}$ is an eigenstate of the undriven ($\eps_d=0$) circuit QED Hamiltonian $\hat H_S$, labeled according to the scheme of \cite{shillito_et_al_2022}. This labeling indicates that, in the limit of vanishing coupling $g=0$, $\ket{\overline{j n}} \to \ket{j} \ket{n}$ with $j=0,1$ denoting qubit state and $n=0,1,\ldots$ the number of photons in the resonator. 

Initialization in a product state with the vacuum of the uncoupled chain $\ket{\mathbf{0}}$ is consistent with the standard treatment of the Lindblad master equation \cite{breuer_petruccione}. Since  \cref{eq:init} is not an eigenstate of the undriven total Hamiltonian \cref{eq:model_chain_2}, under unitary time evolution with $\hat{H}(t)$, entanglement will develop between the system and the bath, even in the undriven case. Our results only minimally change as compared to initializing in the true eigenstates of the full undriven Hamiltonian $\hat{H}$ corresponding to \cref{eq:init} [SM].
Far from the system, the tunneling rates $\{t_i\}$ in \cref{eq:model_chain_2} saturate to a constant value, which fixes a finite speed of propagation of excitations along the chain \cite{chin_chain_mapping}.
This allows us to model dissipative dynamics by means of a coherent exchange of energy between the system and a practically infinite bath, while truncating the chain to a finite number of sites $N$ [SM]. 

To estimate qubit relaxation rates, we consider the following qubit observable expressed in the dressed basis, which takes value $\pm 1$ in the initial state \cref{eq:init}, 
\begin{align}\label{eq:dressed_sigma}
    {\hat \Sigma}_z = \sum_{n=0}^{d-1} \, \dyad{\overline{1n}} - \dyad{\overline{0n}} \; .
\end{align}
It reduces to $\hat\sigma_z$ at vanishing qubit-resonator coupling $g$.
Above, $d$ is the finite Hilbert space dimension for the resonator, which we choose depending on the expected average population of the resonator decoupled from the qubit \cite{Blais_cQED} [SM],
\begin{align}\label{eq:defnbar}
\bar{n} = \eps_d^2 \; / \; [(\omega_a - \omega_d)^2 + \kappa^2/4], \;   
\end{align}
where $\kappa$ is the single-photon decay rate of the readout resonator as introduced in \cref{tab:sdf_table}. We can moreover monitor the readout resonator population in the dressed basis, defined analogously as $\hat N_a = \sum_{n=0}^{d-1} n\left( \dyad{\overline{1n}} + \dyad{\overline{0n}}\right)$.

Additionally, we consider the occupancies of chain normal modes
\begin{align}\label{eq:n_omega}
    \hat n_{\omega_k} = \hat b^\dagger_{\omega_k}\hat b_{\omega_k} \; ,
\end{align}
whose expectation values are related via the chain mapping to two-site correlation functions $\langle c_i^\dagger \hat c_{i'} \rangle$ that are easily accessible in our numerical simulations. Up to the truncation of the chain, the normal modes associated to frequency $\omega_k$ sample $J(\omega)$ [SM]. The operators $\hat{N}_{a}$ and $\hat{\Sigma}_z$ commute with the undriven system Hamiltonian $\hat{H}_S$, whereas $\hat{n}_{\omega_k}$ commutes with the decoupled chain Hamiltonian on the second line of \cref{eq:model_chain_2}. Changes in expectation values of these observables will be proxies for the coherent energy transfer between the drive, the system, and the frequency-resolved modes of the bath.

\paragraph{Results.---} We numerically simulate dispersive readout. For our simulations, we take the following typical values for the system parameters in \cref{eq:model_chain_2}: $\omega_q/2\pi = 5.304\,\text{GHz}$, $\omega_a/2\pi = 7.5\,\text{GHz}$, $g/2\pi = 0.3165\,\text{GHz}$. This places us in the dispersive regime with $|g/\Delta| \simeq 0.14$ with $\Delta = \omega_a - \omega_q$. As mentioned above, we consider several forms of the spectral function $J(\omega)$ --- flat, Ohmic, and Ohmic with a notch filter around the qubit frequency --- while enforcing that the corresponding single-photon decay rate of the resonator $\kappa = 2\pi J(\omega_a) = 50\text{ MHz}$ (see \Cref{tab:sdf_table}). 

Under the system-bath interaction in \cref{eq:model_chain_2}, the system spectral response will be dressed by the bath modes in a way that is dependent on $J(\omega)$. We start with an undriven free decay numerical spectroscopy ($\eps_d=0$) as illustrated on \cref{fig:calibration}, that allows us to determine the effective resonator frequency, which is Lamb shifted by both the qubit and the bath modes. We initialize the system according to \cref{eq:init} with  $\ket{\overline{0n}}$ or $\ket{\overline{1n}}$, and let it evolve until the resonator photon number $\hat N_a$ relaxes to zero (top-right inset of~\cref{fig:calibration}), which in our parameter regime matches the Weisskopf-Wigner theory of spontaneous emission  \cite{Weisskopf_Wigner_1930,Sharafiev2021Jun} [SM]. 
\Cref{fig:calibration} shows the final time occupancies of the chain normal modes $\hat n_{\omega_k}$ of \cref{eq:n_omega}, for initial conditions, with $n=2$ resonator photons to account for a typical value of the ac Stark shift of the resonator in our numerical experiments, and an Ohmic bath spectrum. Each curve exhibits a dominant peak at frequency close to the bare resonator frequency $\omega_a$, and, when the qubit is initially excited, there is an additional peak close to the bare qubit frequency (bottom right inset). The positions of the peaks at frequencies $\omega_{aj}$ close to $\omega_a$ depend on the initial states of the qubit $j=0,1$. Note that $\omega_{aj}$ depend on $J(\omega)$ and we hence repeat this calibration for each bath type. In dispersive readout, the frequency difference between these two peaks, called the dispersive shift, is exploited in order to measure the state of the qubit \cite{blais_et_al_2004}. 

We now turn to the simulation of an actual dispersive readout experiment. We initialize the system according to \cref{eq:init} with $j=1$ excitation in the qubit and $n=0$ photons in the readout resonator. We choose to place the readout drive frequency $\omega_d$ at the midpoint $\bar \omega = (\omega_{a0} + \omega_{a1})/2$, consistent with maximizing the contrast in the phase response of the readout resonator. Other choices for $\omega_d$ are possible, and depending on the parameter regime, can be expected to give quantitatively different results \cite{Huang2026Mar}. We have run numerical simulations for the various spectral functions of \cref{tab:sdf_table} and for multiple drive amplitudes, targeting steady-state resonator photon numbers in the range $0 \leq \bar{n} \lesssim 6.5$. In each of these numerical experiments, fitting an exponential decay to the time dependence of $\bra{\psi(t)} \hat \Sigma_z \ket{\psi(t)}$ (see \cref{fig:spect_sigz} c)-d)), we can extract the qubit energy relaxation rate $\Gamma_{10}(\bar{n})$.

\begin{figure}[t!]
    \centering
    \includegraphics[width=0.95\linewidth]{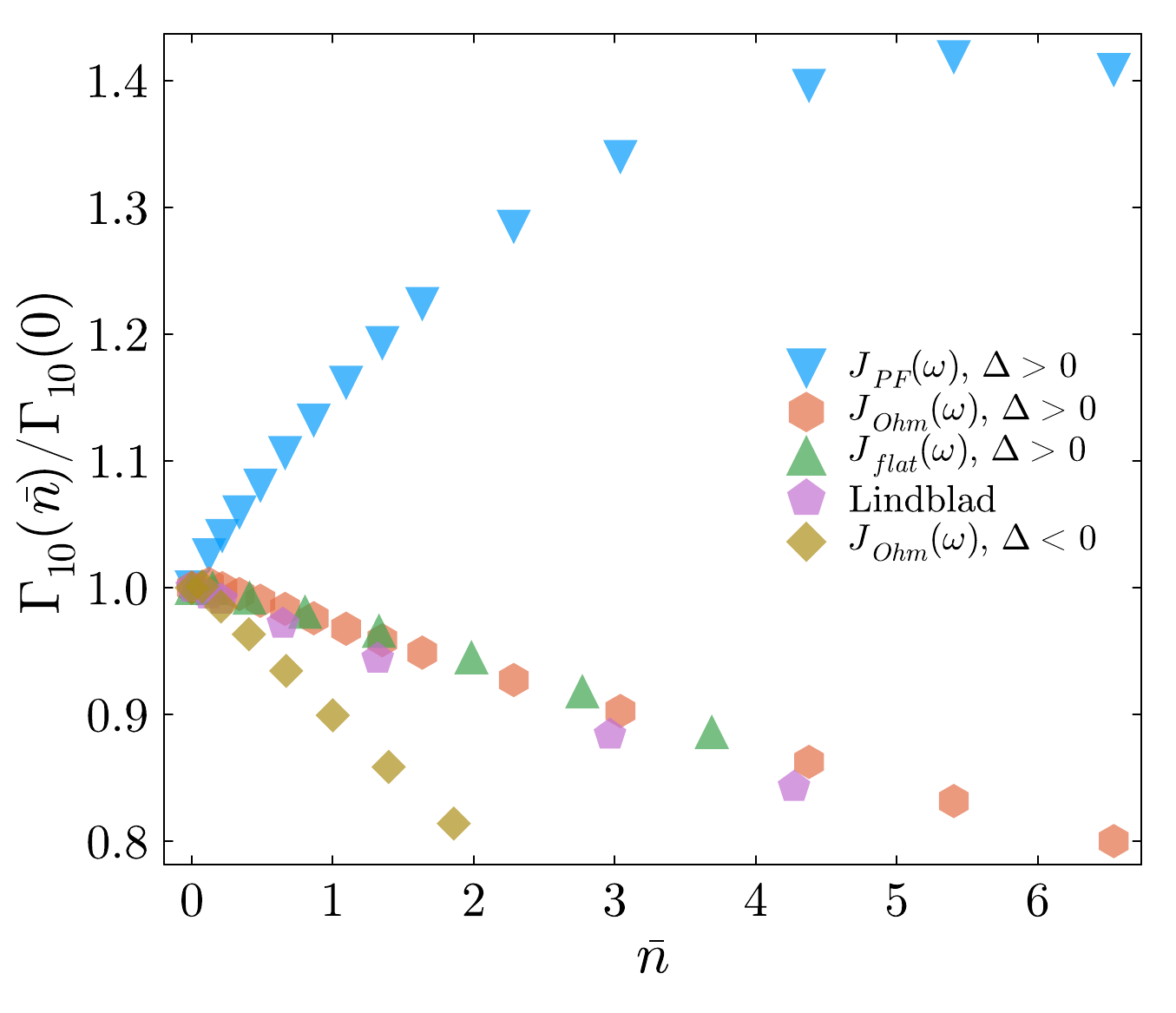}
    \caption{\emph{Inverse $T_1$ versus $\bar n$ characterization (MPS simulation):} The relative change in relaxation rate $\Gamma_{10}(\bar n)/\Gamma_{10}(0)$ of the qubit is shown for various $J(\omega)$, with a comparison to the prediction of the Lindblad master equation. Different spectral densities give rise to distinct relaxation dependence on drive power, with a change in qualitative behavior at increasing drive power for the Purcell filter case, the only case among the ones considered in our study where $T_1$ decreases with $\bar n$. Swapping qubit and resonator frequencies ($\Delta <0$, gold rhombi) gives rise to a starker dependence on $\bar{n}$.}
    \label{fig:t1_vs_nbar}
\end{figure}

As shown in \cref{fig:t1_vs_nbar}, we find that the dependence of $\Gamma_{10}(\bar{n})$ versus $\bar{n}$ is very sensitive to the choice of spectral function for the bath. For Ohmic and flat spectral functions, with the parameters in \cref{tab:sdf_table}, the qubit relaxation rate $\Gamma_{10}(\bar{n})$ decreases with readout power, whereas for our choice of a notch Purcell filter around the qubit frequency (see \cref{tab:sdf_table}), it increases with $\bar{n}$ \footnote{For consistency across the different simulations, we label the $x$-axis with $\bar{n}$ according to the expression under \cref{eq:defnbar}, where the drive frequency $\omega_d$ is set according to the calibrated value $\bar{\omega}$ above, and $\omega_a$ is replaced by $\omega_{a1}$, the resonator frequency as pulled by the qubit excited state.}. For the drive amplitudes considered, we obtain only a negligible contribution to energy relaxation time $T_1$ from the \emph{excitation} rate $\Gamma_{01}(\bar{n})$ upon initializing in $\ket{\overline{00}}$ [SM]. Thus, we show that the energy relaxation time $T_1$ can be degraded or improved as the readout photon number $\bar{n}$ is increased, depending on the fine details of the bath spectral function, which is one of our main results.

We emphasize that an increase of $\Gamma_{10}(\bar{n})$, which we attribute to features in the bath spectral function, cannot be predicted by models commonly used in the previous literature. The Lindblad master equation \begin{align}
    \dot{\rho} = -i[\hat{H}_S(t),\hat\rho] + \kappa \left[\hat{a} \hat \rho \hat a^\dagger - \half \{ \hat a^\dagger  \hat a, \hat \rho \}\right] \label{eq:lindblad}
\end{align}
 predicts a monotonic decrease of the qubit decay rate $\Gamma_{10}$ with $\bar{n}$, consistent with \cite{sete_gambetta}. This qualitatively deviates from our numerical simulations in the case of the Purcell filtered $J_{PF}(\omega)$. Moreover, as we are not making the secular approximation in either the system-to-bath coupling or the coupling between qubit and readout resonator in \cref{eq:model_star}, there are further observations in our converged numerical results that  \cref{eq:lindblad} cannot reproduce [SM]. 
For instance, the Lindblad master equation predicts a spurious excitation rate upon initializing in $\ket{\overline{00}}$, while the MPS simulations account for the correct stationary behavior.

To shed some light on the phenomena at play, we can take advantage of our access to bath 
mode occupations $\langle \hat{n}_{\omega_k} \rangle$ to extract spectroscopic observables for the qubit. In \cref{fig:spect_sigz}a) and b) we show $\langle \hat{n}_\omega \rangle$ as a function of readout photon number $\bar{n}$ for the Ohmic and Purcell filtered Ohmic scenarios. As the drive amplitude is increased, the qubit peak is ac Stark shifted and broadened, as expected from the theory of dispersive readout \cite{Blais_cQED}. For all the studied scenarios, we are able to relate variations in the  monotonicity of the resonator-filtered bath spectral function $\tilde{J}(\omega)$, obtained by grouping the resonator with the bath, and inverting the chain mapping [superposed on \cref{fig:spect_sigz} a)-b)], to the changes in $T_1$. To test this, we have swapped the qubit and resonator frequencies ($\Delta <0$ with $J_\text{Ohm}(\omega)$ in \cref{fig:t1_vs_nbar}). In this case, the qubit is ac Stark shifted to higher frequencies by the readout drive, but the rate $\Gamma_{10}(\bar{n})$ is still decreasing because the qubit is now probing a monotonically decreasing $\tilde{J}_\text{Ohm}(\omega)$ [SM].

This is consistent with the picture of \cite{thorbeck,Huang2026Mar}, where qubit relaxation rates could increase as its resonance is ac Stark shifted and broadened into `hot spots' of a bath directly coupled to it. For our distinct scenario of cavity-mediated relaxation mechanisms, however, there exists no analytical theory, and our numerical results could pave the way towards benchmarking new approaches \cite{cochin2026}. 

\begin{figure}[t!]
    \centering
    \includegraphics[width=1.0\linewidth]{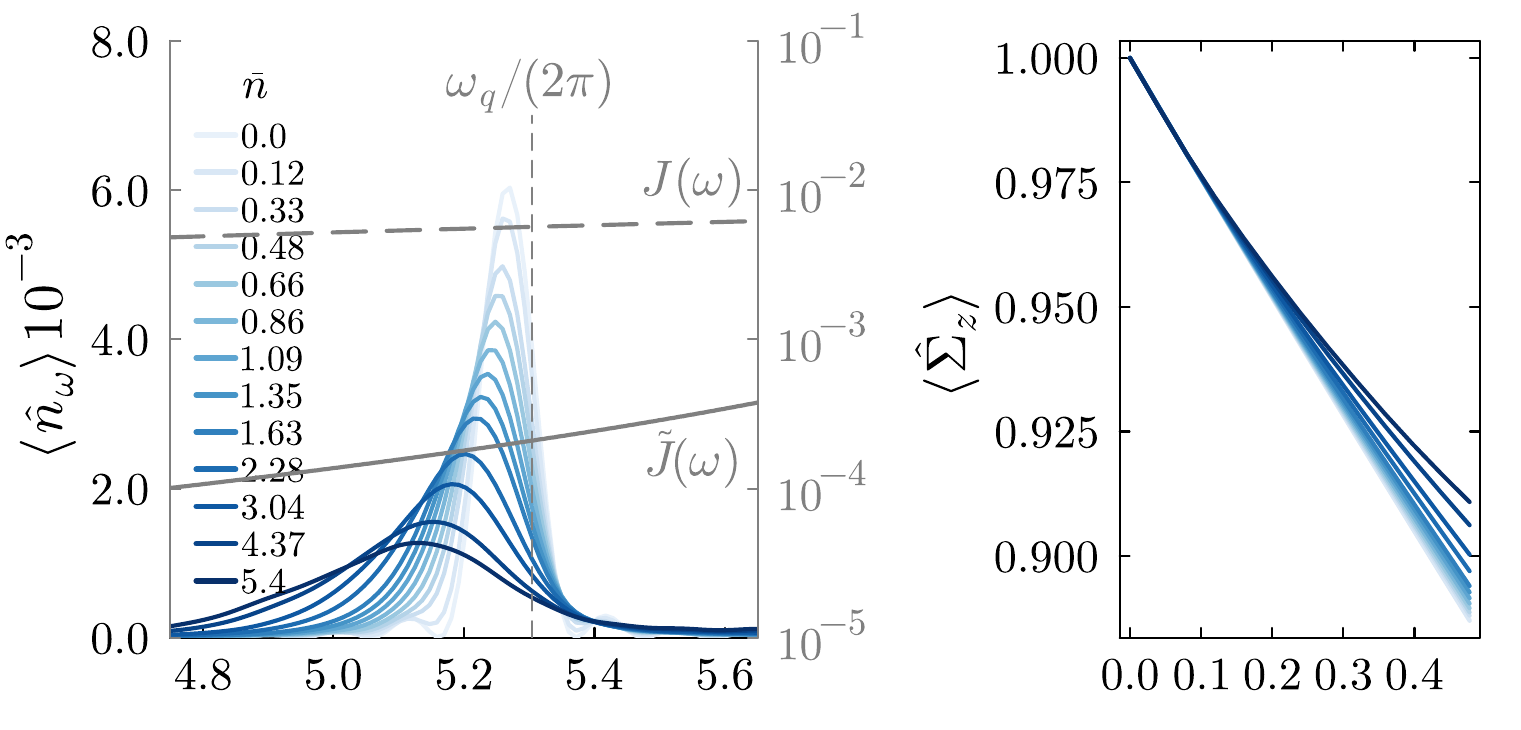} 
    \includegraphics[width=1.0\linewidth]{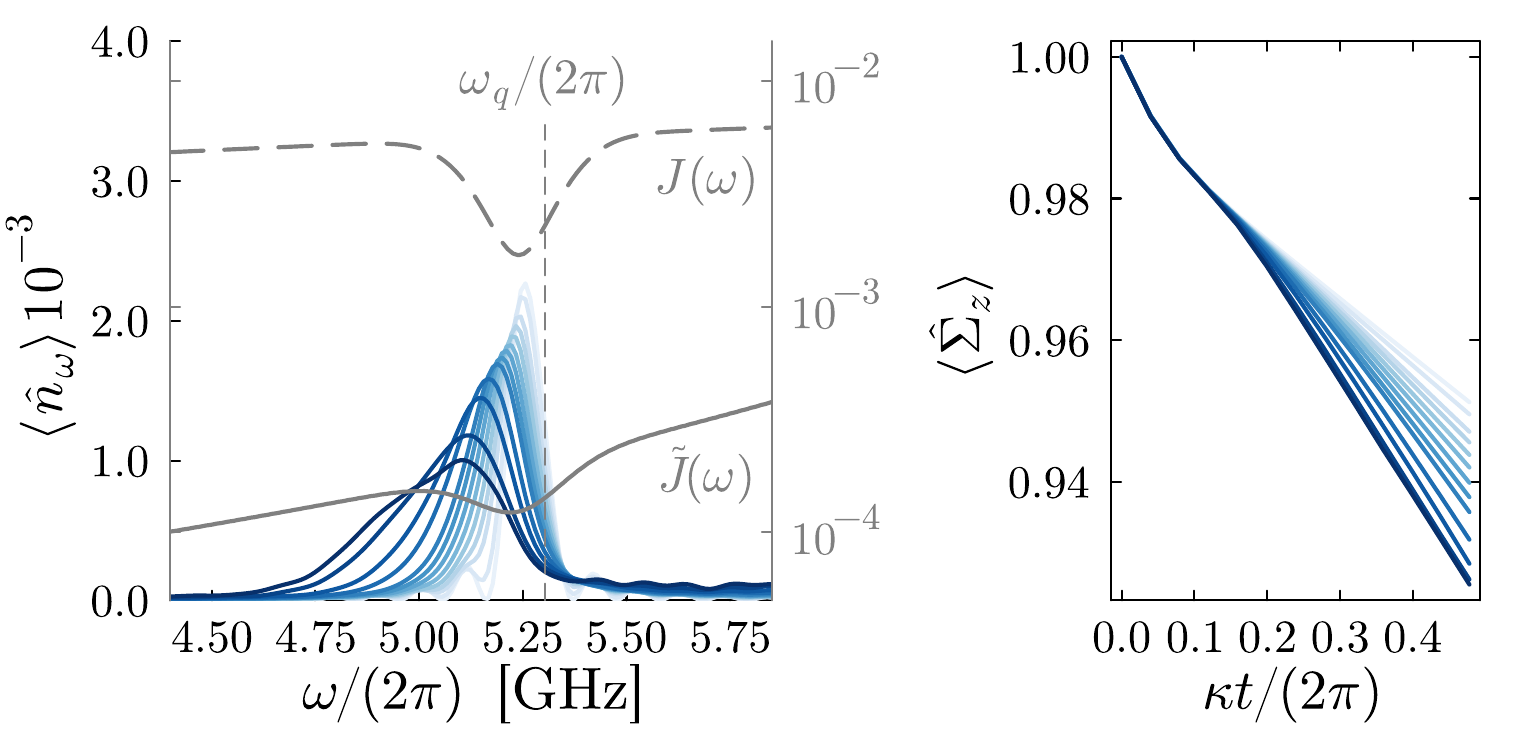}
    \caption{\emph{Qubit spectral analysis (MPS simulation):} Bath mode occupations $\expval{\hat n_\omega}$ at time $\kappa t/(2\pi)= 0.5$, for $\omega$ around the qubit frequency (left, blue curves), for the Ohmic $J_{\rm Ohm}(\omega)$ (top) and Purcell filter $J_{\rm PF}(\omega)$ (bottom) bath spectral densities, and drive power increasing with darker color. These observables reflect the ac Stark shift and broadening of the qubit resonance. We superimpose 
    $J(\omega)$ and the resonator-filtered spectrum $\tilde{J}(\omega)$, as seen by the qubit, on a logarithmic scale in units of GHz. We can correlate an increase in $\Gamma_{10}(\bar{n})$  of \cref{fig:t1_vs_nbar} with the qubit resonance shifting towards higher values of $\tilde{J}(\omega)$, and vice-versa. We extract $\Gamma_{10}(\bar{n})$ from the corresponding decay of qubit $\langle{\hat \Sigma_z}\rangle$, as shown on the right, by fitting an exponential over the interval $0.2 \leq \kappa t/2\pi \leq 0.5$.} 
    \label{fig:spect_sigz}
\end{figure}

\textit{Conclusions.--} The dependence of qubit $T_1$ on photon number $\bar{n}$ in a dispersively coupled readout resonator remains an outstanding problem \cite{Slichter2012Oct,Mundhada2016,Gusenkova2020Sep,Hutin2024Oct,Dai2025Jun,Bista2025Jan,May2025Feb} amounting to a significant portion of the error budget in state-of-the-art quantum processors \cite{Acharya2025Feb}. In this Letter, we have accessed the non-Markovian qubit relaxation by simulating the full time-evolution, using a first-principles Hamiltonian derived from the bath spectral density. To isolate mechanisms independent of measurement-induced leakage involving qubit states outside of the computational manifold, we have restricted the analysis to a two-level system. We have been able to relate features in the bath spectral density to changes in the qubit relaxation as the readout power is increased. We provided insight into this mechanism thanks to access to bath observables, which reveal the  emission spectrum of the circuit QED system. Our analysis can be applied to any experimentally motivated bath spectral function, and used to improve filter design. Future work will focus on its interplay with state transitions outside the computational subspace, taking into account multilevel systems such as transmons or fluxonia.

\textit{Acknowledgments.--} The authors wish to thank Brieuc Le Dé for help with setting up the numerical simulations for driven systems. Simulations were performed using \texttt{MPSDynamics.jl} \cite{mpsdynamicsjl_2024,mpsdynamics_zenodo2021} on the CLEPS cluster at Inria Paris. The authors acknowledge support from ANR project MecaFlux (ANR-21-CE47-0011), ANR project ANR-22-CE30-0033, PCQT (QuanTEdu France) and PEPR integrated project EPiQ ANR-22-PETQ-0007.

\bibliographystyle{apsrev4-2}
\bibliography{biblio}

\onecolumngrid
\newpage

\appendix

\section{SUPPLEMENTAL MATERIAL}

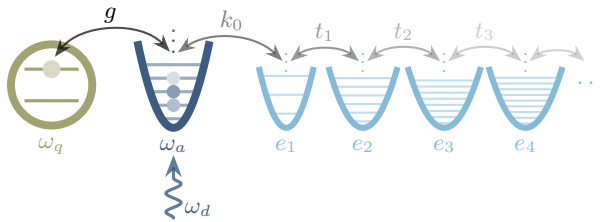
\begin{figure}[b!]
    \centering

    \def\H{1.35}   
\usetikzlibrary{arrows.meta}

\tikzset{
    myarrow/.style={
        thick,
        <->,                
        >=Stealth,          
        shorten <=2pt,
        shorten >=2pt
    }
}

\begin{tikzpicture}[scale=0.6]

    \foreach \x/\y in {-1.1/0.2} {
    \begin{scope}[shift={(\x,\y)}]
        \draw[very thick, kaki_blais, domain=-0.8:0.8, samples=100]  (-2.2, 2) -- (-1., 2);
        \draw[very thick, kaki_blais, domain=-0.8:0.8, samples=100]  (-2.2, 2.7) -- (-1., 2.7);
        \draw[line width=3pt, kaki_blais, domain=-0.8:0.8, samples=100] (-1.6, 2.35) circle (0.9cm);

        \fill[kaki_blais!40] (-1.6, 2.7) circle (0.2cm);
        \node[kaki_blais, ultra thick] at (-1.6, 1.) {$\omega_{q}$};  
        \node[ultra thick] at (-0.3, 3.9) {$g$};  
    \end{scope}
    }

  \draw[line width=3pt, blue_blais, domain=-0.8:0.8, samples=100] plot (\x, {1.6+3*\x*\x});
      \node[blue_blais, ultra thick] at (0, 3.7) {$\vdots$};  
      \node[blue_blais, ultra thick] at (0, 1.2) {$\omega_a$};  
      
  \foreach \y in {1.8, 2.1, 2.4, 2.7, 3} {
    \pgfmathsetmacro{\xlim}{sqrt((\y-1.6)/3)}
    \draw[blue_blais, opacity=0.5, very thick] (-\xlim,\y) -- (\xlim,\y);
  }
  \fill[blue_blais!40] (0, 2.1) circle (0.15cm);
  \fill[blue_blais!60] (0, 2.4) circle (0.15cm);
  \fill[blue_blais!20] (0, 2.7) circle (0.15cm);

  \draw[wigglyarrow] (0,-0.4) -- (0,1);
  \node[blue_blais] at (0.55, -0.2) {$\omega_d$};


  \foreach \x/\y in {2.5/1.6, 4.2/1.6, 6.0/1.6, 7.8/1.6} {
    \begin{scope}[shift={(\x,\y)}]
      \pgfmathsetmacro{\p}{\x}
      \pgfmathsetmacro{\a}{20*e^(-\p)+2}
      \pgfmathsetmacro{\xmax}{sqrt(\H/\a)}

      \draw[azzurro_blais, domain=-\xmax:\xmax, samples=100, line width=2.3pt, opacity=1] plot (\x, {(\a)*\x*\x});
      \node[scale=0.8, azzurro_blais, ultra thick] at (0, 1.59) {$\vdots$};
      
      \foreach \mult in {3, 7, 11} {
          \pgfmathsetmacro{\yconst}{\mult * sqrt(80*e^(-1.2*\p)+0.4)*0.05}
          \pgfmathsetmacro{\xlimr}{sqrt(\yconst/(\a))}
          \draw[azzurro_blais, opacity=0.5, thick] 
            (-\xlimr,\yconst) -- (\xlimr,\yconst);
        }
    \end{scope}
    }

    \foreach \x/\y in {4.2/1.6, 6.0/1.6, 7.8/1.6} {
    \begin{scope}[shift={(\x,\y)}]
      \pgfmathsetmacro{\p}{\x}
      \pgfmathsetmacro{\a}{20*e^(-\p)+2}
      \pgfmathsetmacro{\xmax}{sqrt(\H/\a)}
      
      \foreach \mult in {15, 19, 23} {
          \pgfmathsetmacro{\yconst}{\mult * sqrt(80*e^(-1.2*\p)+0.4)*0.05}
          \pgfmathsetmacro{\xlimr}{sqrt(\yconst/(\a))}
          \draw[azzurro_blais, opacity=0.5, thick] 
            (-\xlimr,\yconst) -- (\xlimr,\yconst);
        }
    \end{scope}
    }

    \foreach \x/\y in {6.0/1.6, 7.8/1.6} {
    \begin{scope}[shift={(\x,\y)}]
      \pgfmathsetmacro{\p}{\x}
      \pgfmathsetmacro{\a}{20*e^(-\p)+2}
      \pgfmathsetmacro{\xmax}{sqrt(\H/\a)}
      
      \foreach \mult in {27, 31} {
          \pgfmathsetmacro{\yconst}{\mult * sqrt(80*e^(-1.2*\p)+0.4)*0.05}
          \pgfmathsetmacro{\xlimr}{sqrt(\yconst/(\a))}
          \draw[azzurro_blais, opacity=0.5, thick] 
            (-\xlimr,\yconst) -- (\xlimr,\yconst);
        }
    \end{scope}
    }
    \foreach \x/\y in {7.8/1.6} {
    \begin{scope}[shift={(\x,\y)}]
      \pgfmathsetmacro{\p}{\x}
      \pgfmathsetmacro{\a}{20*e^(-\p)+2}
      \pgfmathsetmacro{\xmax}{sqrt(\H/\a)}
      
      \foreach \mult in {35} {
          \pgfmathsetmacro{\yconst}{\mult * sqrt(80*e^(-1.2*\p)+0.4)*0.05}
          \pgfmathsetmacro{\xlimr}{sqrt(\yconst/(\a))}
          \draw[azzurro_blais, opacity=0.5, thick] 
            (-\xlimr,\yconst) -- (\xlimr,\yconst);
        }
    \end{scope}
    }

    \coordinate (Q) at (-2.7, 3.08);
    \coordinate (A) at (0, 3.2);
    \coordinate (B1) at (2.5, 3);
    \coordinate (B2) at (4.2, 3);
    \coordinate (B3) at (6, 3);
    \coordinate (B4) at (7.8, 3);
    \coordinate (B5) at (9.2, 3);

    \node[scale=1, azzurro_blais, ultra thick] at (2.5, 1.2) {$e_1$};
    \node[scale=1, azzurro_blais, ultra thick] at (4.2, 1.2) {$e_2$};
    \node[scale=1, azzurro_blais, ultra thick] at (6, 1.2) {$e_3$};
    \node[scale=1, azzurro_blais, ultra thick] at (7.8, 1.2) {$e_4$};


    \draw[myarrow, bend left=50, opacity=0.7] (Q) to (A);
    \draw[myarrow, bend left=40, opacity=0.5] (A) to node[scale=1, pos=0.5, above] {$k_0$} (B1);
    \draw[myarrow, bend left=40, opacity=0.4] (B1) to node[scale=1, pos=0.5, above] {$t_1$} (B2);
    \draw[myarrow, bend left=40, opacity=0.3] (B2) to node[scale=1, pos=0.5, above] {$t_2$} (B3);
    \draw[myarrow, bend left=40, opacity=0.2] (B3) to node[scale=1, pos=0.5, above] {$t_3$} (B4);
    \draw[myarrow, bend left=40, opacity=0.18] (B4) to (B5);

    \node[azzurro_blais, ultra thick] at (9.3, 2.6) {$\dots$};
    
\end{tikzpicture}
        
    \caption{{ \emph{Simulated model:} {As described by  \cref{eq:model_chain_2} in the main text,} the {two-level} qubit of transition frequency $\omega_{q}$ is Rabi-coupled, with strength $g$, to a driven readout resonator mode of frequency $\omega_a$ (drive frequency $\omega_d$). The resonator is coupled to a semi-infinite chain of harmonic oscillators with onsite energies $e_k$ and nearest-neighbor couplings $t_k$, the resonator interacting only with the first site via coupling $k_0$. The chain coefficients encode the bath spectrum $J(\omega)$.}}
    \label{fig:sketch-app}
\end{figure}

This Supplemental Material is organized as follows. Sec.~A outlines how the initial model (\cref{eq:model_star}) is transformed into the equations actually simulated using the chain mapping \cite{chin_chain_mapping}, including the methodology for reconstructing the spectral density function filtered by the resonator $\tilde{J}(\omega)$ and for computing bath observables. In addition, it details the parameters used in the numerical simulations. Sec.~B analyzes the effect of the rotating-wave approximation on the resonator-bath coupling, comparing MPS against Lindblad simulations and Wigner-Weisskopf theory. Sec.~C provides extended driven spectroscopy data for the flat and Ohmic spectral densities. 

\section{A. From microscopic {to simulated} model}\label{app:chain_mapping_qed}

\subsubsection{TEDOPA chain mapping.} The spectral density $J(\omega)$ defines the \textit{chain-mapping} \cite{chin_chain_mapping}, an isometry from the model where the readout mode couples to each bath mode (star-like configuration), to a Hamiltonian where the readout mode only couples to the first mode of a nearest-neighbor coupled, semi-infinite chain of harmonic oscillators. We define new bosonic modes $\hat c_k$, $\hat c_k^\dagger$, with $[\hat c_k, \hat c_l^\dagger] = \delta_{kl}$ by
\begin{align}
\hat c_n^\dagger=\int_0^\infty d\omega\,U_n(\omega)\,\hat b_\omega^\dagger,\qquad
U_n(\omega)=\sqrt{J(\omega)}\,\tilde p_n(\omega),
\end{align}
where $\tilde p_n$ are orthonormal polynomials with respect to the measure \(d\mu(\omega)=J(\omega)\,d\omega\).
This unitary transformation recasts the linear system-boson Hamiltonian of \cref{eq:model_star} into the semi-infinite nearest-neighbor chain Hamiltonian of \cref{eq:model_chain_2}, {and illustrated on \cref{fig:sketch-app},}
with
\begin{align}\label{eq:kap0}
k_0^2=\int_0^\infty J(\omega)\,d\omega,\qquad
e_1=\frac{1}{k_0^2}\int_0^\infty \omega\,J(\omega)\,d\omega,
\end{align}
and the rest of the chain coefficients \(\{e_n,t_n\}_{n> 0}\) given by the three-term recurrence of $\tilde p_n$. While for the flat and Ohmic spectral densities the chain coefficients have a simple analytical form \cite{chin_chain_mapping}, for the Purcell case (see Table~\ref{tab:sdf_table}) we compute them numerically with the routines available at \cite{mpsdynamics_zenodo2021, mpsdynamicsjl_2024}. 

The one-dimensional tight-binding chain of bosonic modes with nearest-neighbor coupling can benefit from the application of numerically efficient tools based on the matrix-product state (MPS) representation of the many-body wavefunction, avoiding the exponential growth of the simulated state space with the number of modes. In particular, it is well suited to the application of tensor-network time evolution methods like TDVP for MPS \cite{SCHOLLWOCK201196, Haegeman_2011}.

\subsubsection{Bath spectrum considerations.}

\paragraph{{Bath spectrum} discretization.--} To simulate the time evolution generated by the semi-infinite tight binding chain of \cref{eq:model_chain_2} we truncate the chain to length $N$. In the original
(star) picture this corresponds to replacing the continuum by $N$ effective
modes at (generally nonuniform) frequencies $\omega_k$ and with couplings
$g_k$. The associated spectral density is then, approximately
\begin{align}\label{eq:sdf_discrete_to_cont}
J(\omega)\;\approx\;\sum_{k=1}^N |g_k|^2\,\delta(\omega-\omega_k),
\end{align}
where the frequencies are not equally spaced and '$\approx$' is in the quadrature (weak) sense: for any smooth function $f$ supported on the sampled band, we have that $\int_0^{+\infty}f(\omega)J(\omega)d\omega = \sum_{k=1}^N |g_k|^2 f(\omega_k) + O(1/N^2)$, i.e. the discrete measure approximates 
$J(\omega)$ under integration. 
\newline

\paragraph{Effective spectral density function $\tilde{J}(\omega)$.--} 
Conversely, given a set of arbitrary chain coefficients $\{e_k, \, t_k, \, k_0\}$, it is possible to reconstruct the corresponding spectral density function, by following the steps below. 
\begin{enumerate}
    \item Diagonalize the $N \times N$ Jacobi matrix of the chain coefficients, the second line of \cref{eq:model_chain_2}, obtaining the eigenvalues $\{\omega_k\}$ and corresponding eigenvectors $\{v^{(k)}\}$.
    \item Given $k_0^2$ defined in \cref{eq:kap0}, we compute the weights (with $g_k \equiv \sqrt{w_k}$) as
    \begin{align}
        w_k = k_0^2 \abs{v_1^{(k)}}^2,
    \end{align}
    where $v_1^{(k)}$ is the first element of the $k-$th eigenvector.
    \item To obtain a smoother function, we convolve the discretized spectral density of \cref{eq:sdf_discrete_to_cont} with a Gaussian or Lorentzian kernel $f_\eta$, where $\eta$ gives the broadening,
    \begin{align}
        J_{\rm smooth} (\omega) = (J * f_\eta)(\omega) = \sum_k w_k f_\eta (\omega - \omega_k).
    \end{align}
    The broadening parameter $\eta$ has to be chosen to be slightly larger than the interval between the sampled frequencies $\omega_k$ to ensure a smooth profile without washing out the physical features of the spectral density.
\end{enumerate}
{We mainly use this procedure} to compute the effective spectral density function, {as seen by the qubit after being} filtered by the {readout resonator}. {Indeed,} since the readout resonator is just another harmonic oscillator, we can consider it as part of a chain defined by the chain coefficients
\begin{align}
    \{\; [\omega_a, e_0, e_1, \, ...], \quad [k_0, t_0, t_1, \, ...], \quad g\;\} \; ,
\end{align}
and reconstruct the corresponding spectral density, which we denote  
$\tilde{J}(\omega)$. Note that this procedure assumes a Jaynes-Cummings like coupling between modes throughout the chain, while in our case the very first coupling, between resonator and first chain site, is actually dipolar [\cref{eq:model_chain_2}]. The resulting approximation only impacts our computation of $\tilde{J}(\omega)$, which we show for its qualitative features.
 Beside this qualitative analysis, in all of our numerical simulations we consider the full dipolar coupling in \cref{eq:model_chain_2}.
\\

\paragraph{Reorganization energy counterterm.--} Following the Caldeira–Leggett construction \cite{CALDEIRA1983374}, the linear quadrature resonator-bath coupling of \cref{eq:model_star} induces a static renormalization of the resonator potential. This is best understood by considering the bath in \cref{eq:model_star} as a discrete set of harmonic oscillators whose equilibrium positions are shifted by the resonator coordinate $\hat{x}{=(\hat a+\hat a^\dagger)/2}$:
\begin{align}
    \sum_k \frac{\hat{p}_k^2}{2m_k} + \frac{1}{2} m_k \omega_k^2 \left( \hat{q}_k - \frac{c_k}{m_k \omega_k^2} \hat{x} \right)^2 =\int d\omega [ \sqrt{J(\omega)} (\hat a + \hat a^\dagger) (\hat b_\omega + \hat b_\omega^\dagger) +  \omega \hat{b}_\omega^\dagger \hat{b}_\omega ] + \lambda(\hat a+\hat a ^\dag)^2.
\end{align}
This yields the reorganization energy term $\lambda(\hat a+\hat a^\dagger)^2$ added to $\hat{H}_S(t)$ of \cref{eq:model_star}, where $\lambda=\int_0^\infty \frac{J(\omega)}{\omega} d\omega$. \newline

\paragraph{Spectroscopy of the chain Hamiltonian.--} In numerical simulations, we track one- and two-site observables in the chain basis of \cref{eq:model_chain_2},
e.g.\ \(\langle \hat c_i^\dagger \hat c_j\rangle\). To recover frequency-resolved observables in the physical (star) basis, we need to invert the chain mapping.
Let \(U\in\mathbb{C}^{N\times N}\) be the unitary that maps chain to star modes {in the discretized model},
\begin{align}
\hat b_i=\sum_{n=1}^N U_{in}\,\hat c_n,\qquad
U^\dagger U=UU^\dagger=\mathbb{I}.
\end{align}
Then the star-basis one-body correlator matrix is
\begin{align}\label{eq:chain_mapping_inversion}
\mathbf{C}^{(b)} \equiv \bigl[\langle \hat b_i^\dagger \hat b_j\rangle\bigr]
= U\,\mathbf{C}^{(c)}\,U^\dagger,\qquad
\mathbf{C}^{(c)} \equiv \bigl[\langle \hat c_n^\dagger \hat c_m\rangle\bigr].
\end{align}
In particular, the occupation of the $i$-th sampled bath frequency, as given by \cref{eq:n_omega}, can be computed as a linear combination of two-site correlators in the chain model.\\

\paragraph{Increasing the bath spectral sampling.-- } For strictly positive $J(\omega)$ defined on a compact support $[0, \omega_c]$, the chain coefficients converge to asymptotic values $e_n\to e_{{\infty}} = \omega_c/2$, $t_n\to t_{{\infty}} = \omega_c/4$. 
Therefore, once the chain coefficients reach convergence, excitations propagate along the chain with a finite group velocity set by the asymptotic hopping. For times $t$ such that the wavefront has not reached the boundary, only the first $\ell(t)\ll N$ sites are populated.
We exploit this fact to increase frequency resolution {without} simulating a longer chain:
(i) compute the chain coefficients up to a larger size \(M>N\) to obtain the denser quadrature \(\{\omega_k^{(M)},w_k^{(M)}\}\) {of the bath spectral density $J(\omega)$,} and the corresponding unitary \(U^{(M)}\);
(ii) embed the measured \(\mathbf{C}^{(c)}_{N\times N}\) into an \(M\times M\) matrix {$\mathbf{C}^{(c,{\rm pad})}$} 
by zero-padding the rows/columns corresponding to unpopulated sites;
(iii) form \(\mathbf{C}^{(b,M)}=U^{(M)}\,\mathbf{C}^{(c,{\rm pad})}\,(U^{(M)})^\dagger\).
This yields frequency-resolved observables on a finer grid at the cost of building \(U^{(M)}\) {for post-processing}, but \emph{without} increasing the MPS size {of the simulation}, provided reflections from the truncated tail are negligible over the time window of interest \cite{Riva_23}.\\

\subsubsection{{Parameters for the numerical simulations.}}

{Using the above bath model, we have performed numerical simulations of the full time evolution, without any rotating wave approximations. The further approximations made to obtain a tractable model are: (i) finite sampling time $dt$; (ii) finite local dimension, truncating the resonators; and (iii) finite bond dimension in the MPS model, effectively truncating the entanglement or correlation among chain sites. The latter is instrumental in avoiding exponential increase of the simulated state space with chain length, and it is known to provide excellent results for weakly excited one-dimensional chains.}

\begin{table}[]
    \centering
    \renewcommand{\arraystretch}{1.7}
    \begin{tabular}{c|c|c|c}

         & $\, \, \chi \, \, $  & $\, \, \kappa {dt} /(2 \pi) \, \, $ & $\, \, N \, \, $ \\
         
         \hline
       
       $\, \, J_{\rm Ohm}(\omega)\, \, $  &   $\, \, 6\, \, $  &  $\, \, 4\times 10^{-5} \, \, $  & $\, \, 471\, \, $ \\
       $\, \, J_{\rm flat}(\omega)\, \, $  &  $\, \, 5\, \, $   &  $\, \, 4\times 10^{-5} \, \, $  & $\, \, 289\, \, $ \\
       $\, \, J_{\rm PF}(\omega)\, \, $  &   $\, \, 8\, \, $  &  $\, \, 4\times 10^{-5}\, \, $  & $\, \, 472\, \, $ \\
    \end{tabular}
    \caption{Simulation parameters for the MPS time evolution with TDVP: bond dimension $\chi$, dimensionless time step $\kappa {dt} /(2 \pi)$, chain length $N$.}
    \label{tab:conv}
\end{table}

The parameter values used for our numerical simulations up to the final time $\kappa t / (2\pi)= 0.5$ are reported in \cref{tab:conv}. {We next give some details on how they were selected.}\\

\paragraph{Time evolution.--} Numerical simulations were performed using the Julia package \texttt{MPSDynamics.jl} \cite{mpsdynamics_zenodo2021,mpsdynamicsjl_2024}. In particular, we compute the time evolution using the one-site time dependent variational principle algorithm for matrix product states \cite{Haegeman_2011}, which constrains the evolution on a manifold in the Hilbert space characterized by those states that can be represented as matrix product states of a fixed bond dimension $\chi$, set at the beginning of the simulation. Constraining the dynamics on a manifold introduces a projection error, sensitive to both the bond dimension $\chi$ defining the manifold, and the discretization time step $dt$. We computed the entanglement entropy of the MPS at multiple time frames, and the time evolution of $\hat \Sigma_z(t)$ and $\hat N_a(t)$ for all of the configurations considered. {We have simulated with various} values of both $dt$ and $\chi$ {to ensure that the simulations have converged.}\\

\paragraph{Local dimensions.--} The dimensions of the local Hilbert space of each harmonic oscillator has been truncated {in the following way.} 
For the {readout resonator}, which is resonantly driven at amplitude $\eps_d$, we set the local Hilbert space dimension as
\begin{align}
\label{eq:dR_rule}
d_a \;=\; \Big\lceil\, 10 \;+\; 7\sqrt{\bar n_a}\;+\;\bar n_a\,\Big\rceil,
\end{align}
where $\bar n_a = 4 \eps_d^2/\kappa^2$ {is a bath-spectrum-independent approximation to \cref{eq:defnbar} in the main text}.
For the bosonic chain sites, which are not directly driven, we estimate the average occupation of the first site  from the chain coefficients of~\cref{eq:kap0} as $\bar n_{\rm chain} \;=\; \frac{t_0}{k_0}\,\bar n_a$,
where $k_0$ is the {resonator-to}–chain coupling and $t_0$ the hopping between the first and second chain sites. We then choose a uniform Fock cutoff for all chain sites
\begin{align}
d_{\rm chain} \;=\; \Big\lceil\, 2 \;+\; 5\,\bar n_{\rm chain}\,\Big\rceil,
\end{align}
where the constants were chosen after some convergence tests. In particular, we monitor the error
\begin{align}\label{eq:delta_err}
    \delta_{\rm sat}(t) = 1- \expval{[\hat a, \, \hat a^\dagger] (t)},
\end{align}
which could deviate from zero due to the finite truncation $d_a$ of the Hilbert space. We confirmed numerically that the resonator and chain Hilbert-space truncation does not bias our results by checking that
$\delta_{\rm sat}(t)$ is always lower than $10^{-6}$.

\begin{figure}[t!]
    \centering
    \includegraphics[width=0.5\linewidth]{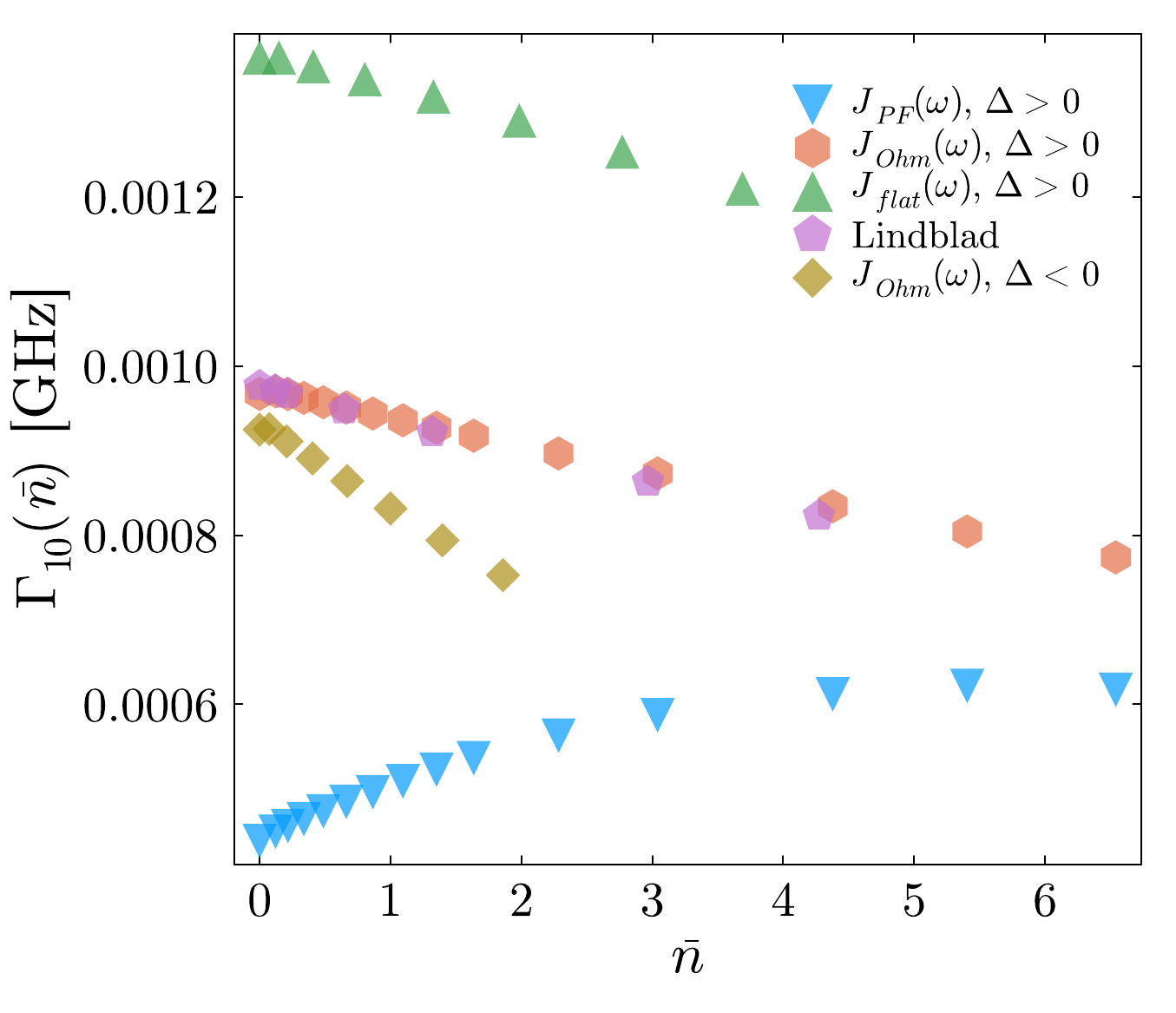}
    \caption{Inverse $T_1$ vs. $\bar n$. The absolute change in relaxation rate $\Gamma_{10}(\bar n)$ is shown for multiple $J(\omega)$, with a comparison to Lindblad master equation. Different spectral densities give rise to distinct relaxation dependence of the qubit on drive power, with a change in qualitative behavior at increasing drive power for the Purcell filter case, the only case among the ones here considered where $T_1$ decreases with $\bar n$. At $\bar n = 0$, note the effect of the Purcell filter in suppressing $\Gamma_{10}(\bar n)$, all other model parameters, including the value of the spectral density function at the resonator frequency $J(\omega_a)=\kappa/(2\pi)$, held equal.}
    \label{fig:t1_vs_nbar_abs}
\end{figure}

\begin{figure}
    \centering
    \includegraphics[width=0.5\linewidth]{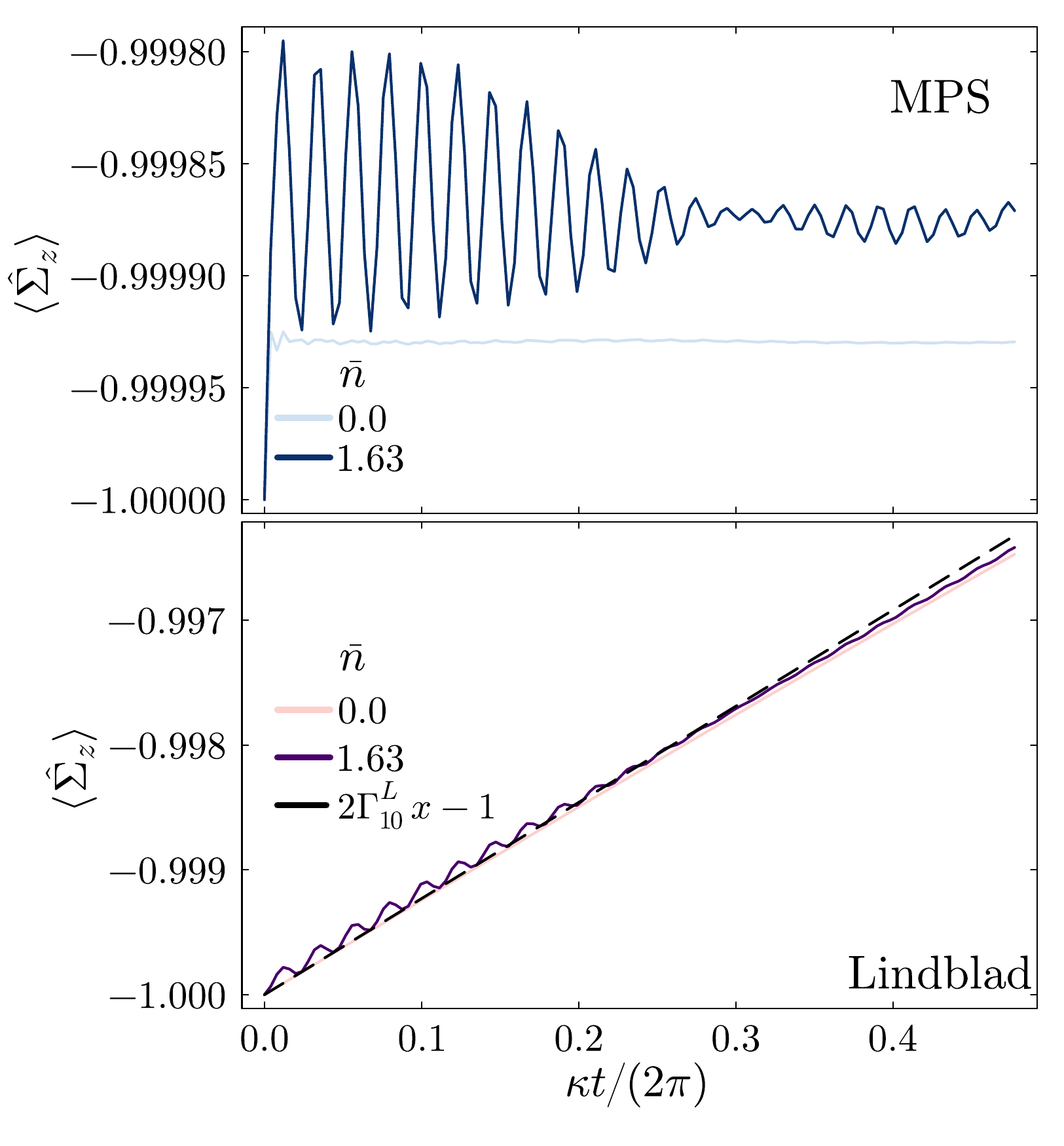}
    \caption{Time evolution of $\langle{\hat \Sigma_z}\rangle$, for the undriven case $\bar n=0$ and for the driven case with $\bar n=1.63$, with initial state $\ket{\overline{00}}\otimes\ket{\mathbf{}{0}}$. The top panel shows the MPS simulation results, while the bottom panel shows the Lindblad results.}
    \label{fig:ground_state}
\end{figure}

\section{B. Analytic differences between dipolar bath coupling and Lindblad model}

Our numerical study considers the full dipolar coupling between system and bath and the full time-dependent Schr\"odinger evolution of the system and bath according to \cref{eq:model_star} and \cref{eq:model_chain_2}. In this section, we highlight the differences between a standard determination of rates from the full microscopic model \cref{eq:model_star}, and those obtained from the Lindblad model \cref{eq:lindblad}.\\

\paragraph{Evaluating the absolute decay rates.--} Following \cite{Beaudoin_2011}, we identify perturbatively the first eigenstates of the Rabi Hamiltonian $\hat H_S = \frac{\omega_{q}}{2} \hat \sigma_z + \omega_a \hat a^\dagger \hat a + g \hat \sigma_x(\hat a+\hat a^\dagger)$, namely
\begin{align}
\label{eq:eigenstates}
    \begin{split}
        &\ket{\overline{00}} = \ket{00} - \frac{g}{\Sigma}\ket{11} + \mathcal{O}\Big(\abs{\frac{g}{\Delta}}^2\Big)\\
        &\ket{\overline{10}} = \ket{10} - \frac{g}{\Delta} \ket{01}+\mathcal{O}\Big(\abs{\frac{g}{\Delta}}^2\Big)
    \end{split}
\end{align}
where the term proportional to $\Sigma = \omega_a + \omega_q$ is due to the counter-rotating terms $\hat\sigma^- \hat a$ and $\hat\sigma^+ \hat a^\dagger$ of the Rabi model. In particular, even for higher order terms, the parity of the total number of excitations in the system, $e^{i \pi (\hat{\sigma}^+\hat{\sigma}^- + \hat{a}^\dag \hat{a})}$, is the same on the left and right side of \cref{eq:eigenstates}.

The Lindblad \cref{eq:lindblad}, at zero readout drive amplitude, evaluated in the hybridized eigenbasis of the Rabi Hamiltonian $\hat{H}_S$, predicts a qubit decay rate at zero photon number. This is due to the single-photon decay term $\kappa \mathcal{D}[\hat{\rho}]$ and reads
\begin{align}
\label{app:eq:LBdecay}
    \Gamma_{10}^{\rm L}(0) \approx 
    \kappa\; \bra{\overline{00}} \; \left(\hat a\ket{\overline{10}}\bra{\overline{10}}\hat{a}^\dagger- \tfrac{\hat{a}^\dagger \hat{a}\ket{\overline{10}}\bra{\overline{10}}}{2} - \tfrac{\ket{\overline{10}}\bra{\overline{10}}\hat{a}^\dagger \hat{a}}{2}\right)\; \ket{\overline{00}} = \kappa \frac{g^2}{\Delta^2}, \;
\end{align}
which is the known form for the Purcell decay rate in this regime \cite{boissonneault_et_al_2009,sete_gambetta}. 

To estimate the rate we would obtain by performing the standard Born, Markov, and secular approximations to obtain a Lindblad master equation from the microscopic model \cref{eq:model_star}, see e.g. \cite{breuer_petruccione}, we need to inspect the system-bath coupling $(\hat a + \hat a^\dagger)\otimes \hat B$, where $\hat B$ is the interaction operator on the bath side, and to apply Fermi's Golden Rule evaluated between the dressed
eigenstates of the Rabi Hamiltonian. This yields the undriven decay rate
\begin{align}\label{app:eq:WCRdecay}
    \Gamma_{10}(0) \approx 2\pi J(E_{\overline{10}} - E_{\overline{00}}) \abs{\bra{\overline{00}}\hat a + \hat a^\dagger\ket{\overline{10}}}^2 = 2\pi J(\omega_q) {g^2}\abs{\frac{1}{\Delta} + \frac{1}{\Sigma}}^2.
\end{align}
The zero-drive rates $\Gamma_{10}(0)$ computed in simulations, and shown in \cref{fig:t1_vs_nbar_abs}, are in reasonable agreement with the analytical predictions of \cref{app:eq:WCRdecay}. For example, the analytical estimate for the undriven decay rate $\Gamma_{10}(0)/(2\pi) \approx 1.43$ MHz for a flat spectral density $J_{\rm flat}(\omega)$ is in close agreement with the value of $1.36$ MHz obtained from MPS simulations (see \cref{fig:t1_vs_nbar_abs}). This value significantly differs from the $1.0$ MHz obtained from the Lindblad model \eqref{app:eq:LBdecay}. 

The numerical difference in the previous paragraph can be understood as follows. The microscopic-model-derived rate \cref{app:eq:WCRdecay} depends on the power spectral density evaluated at the qubit frequency. The disagreement with \cref{app:eq:LBdecay}, which is derived under the same flat spectrum assumption $2\pi J(\omega_q)=2\pi J(\omega_a) = \kappa$, is due to the contributions from the Rabi counter-rotating terms (in $1/\Sigma$) in \cref{app:eq:WCRdecay}. For other bath spectral densities, since \eqref{app:eq:WCRdecay} requires evaluating the spectral function at $J(\omega_q) \neq J(\omega_a) = \kappa/(2\pi)$, the difference between the two rates is further increased. In particular, the Purcell filter efficiently suppresses the qubit decay by design. Note that the treatment here does not hold in the driven case: the frequency of the qubit is ac Stark shifted, the drive further dresses the states, and the resonator gets populated at $\bar n$ higher than zero. 

\paragraph{Evaluating the qubit excitation rate.--} Starting from the ground state of the hybridized eigenbasis of the Rabi Hamiltonian, the Lindblad model would predict a qubit excitation rate
\begin{align}\label{app:eq:LBexcite}
\Gamma_{01}^{\rm L}(0) \approx 
    \kappa\; \bra{\overline{10}} \; \left(\hat a\ket{\overline{00}}\bra{\overline{00}}\hat{a}^\dagger- \tfrac{\hat{a}^\dagger \hat{a}\ket{\overline{00}}\bra{\overline{00}}}{2} - \tfrac{\ket{\overline{00}}\bra{\overline{00}}\hat{a}^\dagger \hat{a}}{2}\right)\; \ket{\overline{10}} = \kappa \frac{g^2}{\Sigma^2},
\end{align}
while applying the Fermi Golden Rule to the system-bath coupling would imply
\begin{align}
    \Gamma_{01}(0) \approx 2\pi J(E_{\overline{00}} - E_{\overline{10}}) \abs{\bra{\overline{10}}\hat a + \hat a^\dagger\ket{\overline{00}}}^2 = 2\pi J(-\omega_q) {g^2}\abs{\frac{1}{\Delta} + \frac{1}{\Sigma}}^2 = 0,
\end{align}
since $J(\omega)$ is identically zero for $\omega<0$ at zero temperature as considered throughout this study.

In Fig.~\ref{fig:ground_state}, we show the evolution of the $\Sigma_z$ observable for the MPS and Lindblad case, starting from $\ket{\overline{00}}$. While the MPS correctly predicts the absence of a steady-state excitation rate, the short-time dynamics exhibits small coherent oscillations. These are not representative of a physical excitation process, but rather a numerical quench artefact, since the system is initialized in the dressed ground state of the Rabi Hamiltonian $\ket{\overline{00}}$ according to \cref{eq:init}, ignoring the initial entanglement with the chain. On the other hand, the Lindblad results in Fig.~\ref{fig:ground_state} show an unphysical excitation, as expected from~\cref{app:eq:LBexcite}. An effective Lindblad master equation for the qubit, with dissipators derived from the two rates \cref{app:eq:LBdecay} and \cref{app:eq:LBexcite}, $\Gamma_{10}^{\rm L}\, \mathcal{D}[\hat{\sigma}_-] + \Gamma_{01}^{\rm L}\, \mathcal{D}[\hat{\sigma}_+]$, would predict the evolution $\langle \hat\Sigma_z(t) \rangle = -1 + 2\Gamma_{01}^{\rm L} \, t$ at short times when starting with $\langle \hat\Sigma_z(t) \rangle=-1$. This is in good agreement with the observed rate in the coupled qubit-resonator system with Lindbladian dissipation $\kappa \mathcal{D}[\hat{a}]$. Importantly, the negligible excitation rate observed on the upper plot of Fig.~\ref{fig:ground_state} allows us to restrict attention to $\Gamma_{10}$ for the evaluation of $T_1$ in the main text. \\

\begin{figure}
    \centering
    \includegraphics[width=0.5\linewidth]{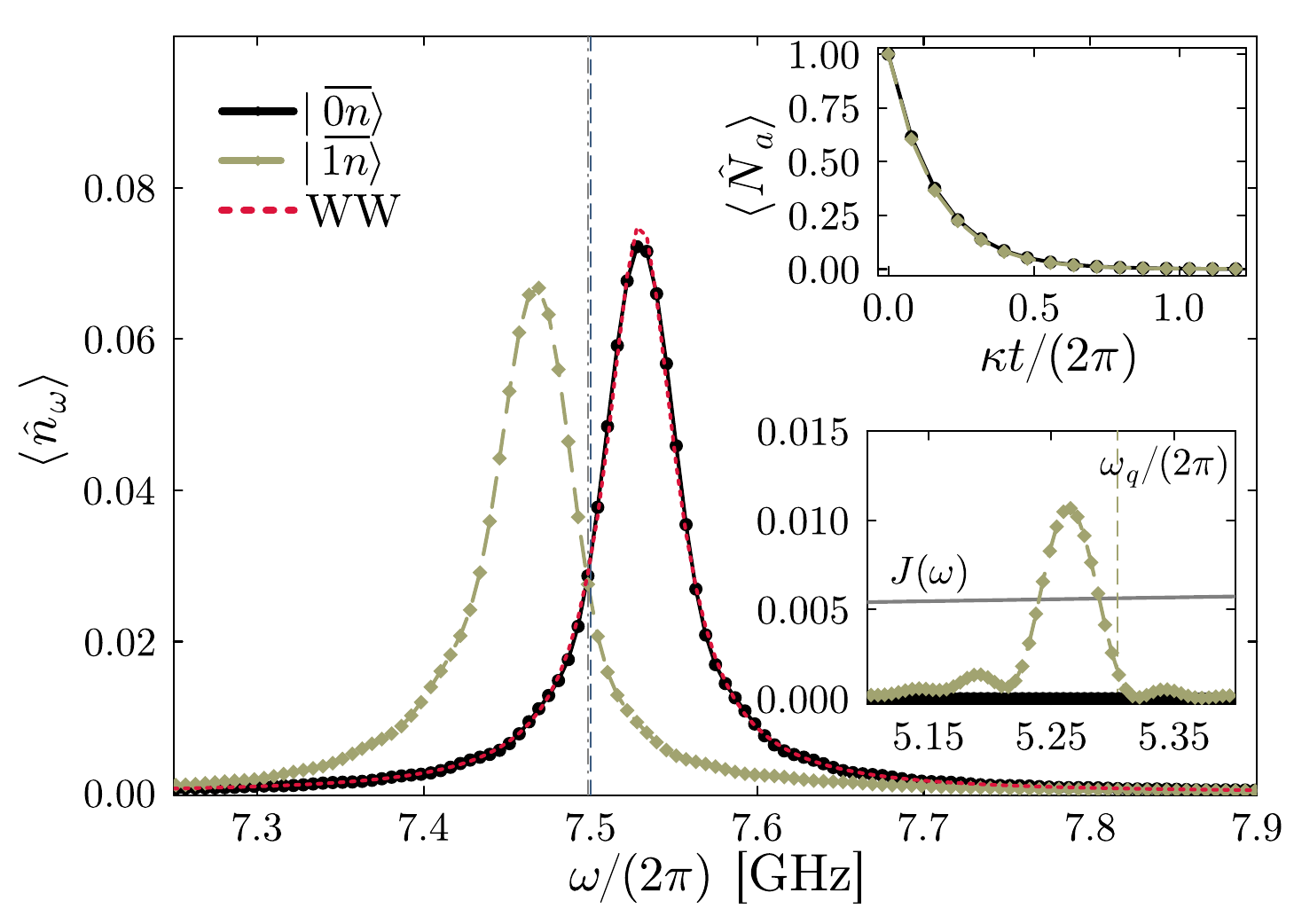}
    \caption{Frequency resolved occupancies $\langle \hat n_\omega \rangle$ of \cref{eq:n_omega}, for the parameters given in the text, with $n=1$ and an Ohmic spectral density function $J_{\rm ohm}(\omega)$. For both initial conditions $\ket{\overline{01}}$ and $\ket{\overline{11}}$, we first let $\langle \hat N_a \rangle$ relax to time $\kappa t/(2\pi)=1$ (top-right inset). We then plot $\langle \hat n_\omega \rangle$ at final time. The Wigner-Weisskopf distribution of \cref{eq:WW} is the dashed red line, which well predicts $\langle \hat n_\omega \rangle$ with $\ket{\overline{01}}$ initial condition.}
    \label{fig:wigner-weisskopf}
\end{figure}

\paragraph{Wigner-Weisskopf theory.--}
The Wigner-Weisskopf theory of spontaneous emission can be applied to the case of the Rabi model \cite{Weisskopf_Wigner_1930, Sharafiev2021Jun}, and it is valid within the resonator-bath secular approximation, i.e.~assuming $\hat H^{\rm RWA} = \hat H_S + \sum_k [ g_k (\hat a^\dagger\hat b_k + \hat a\hat b_k^\dagger) +  \omega_k \hat{b}_k^\dagger \hat{b}_k ]$, with $\hat H_S$ defined above \cref{eq:eigenstates}. Upon initializing the system in \cref{eq:init} with the qubit in its ground state and one photon in the resonator, corresponding to dressed state $\ket{\overline{01}}$,
the expected steady-state occupation $\langle \hat n_{\omega_k}\rangle$ of a bath mode at frequency $\omega_k$ is thus given by the probability distribution
\begin{align}\label{eq:WW}
     \langle \hat n_{\omega_k} \rangle = \frac{J(\omega_k) \Delta\omega_k}{(\kappa/2)^2 + (\omega_k - \omega_{a0} )^2}  \; ,
\end{align}
where the spectral density function $J(\omega_k)$ evaluated at $\omega_k$ is weighted by the corresponding interval of frequency $\Delta\omega_k$, with $\Delta \omega_k = (\omega_{k+1} - \omega_{k-1})/2$. The Lorentzian distribution is centered at the shifted peak frequency $\omega_{a0}$, which accounts for both the dispersive shift induced by the qubit-resonator interaction and the Lamb shift induced by the bath modes. The measured mode occupation curve $\langle \hat n_\omega \rangle$ (black line) in \cref{fig:wigner-weisskopf} shows excellent agreement with the parameter-free theoretical prediction \cref{eq:WW} (red dotted line). The decay rate determined as $\kappa = 2\pi J(\omega_a)$ sets the width of the Lorentzian and is selected to be the same for all the models considered (see \cref{tab:sdf_table}).\\

\begin{figure}[t!]
    \centering
    \includegraphics[width=0.5\linewidth]{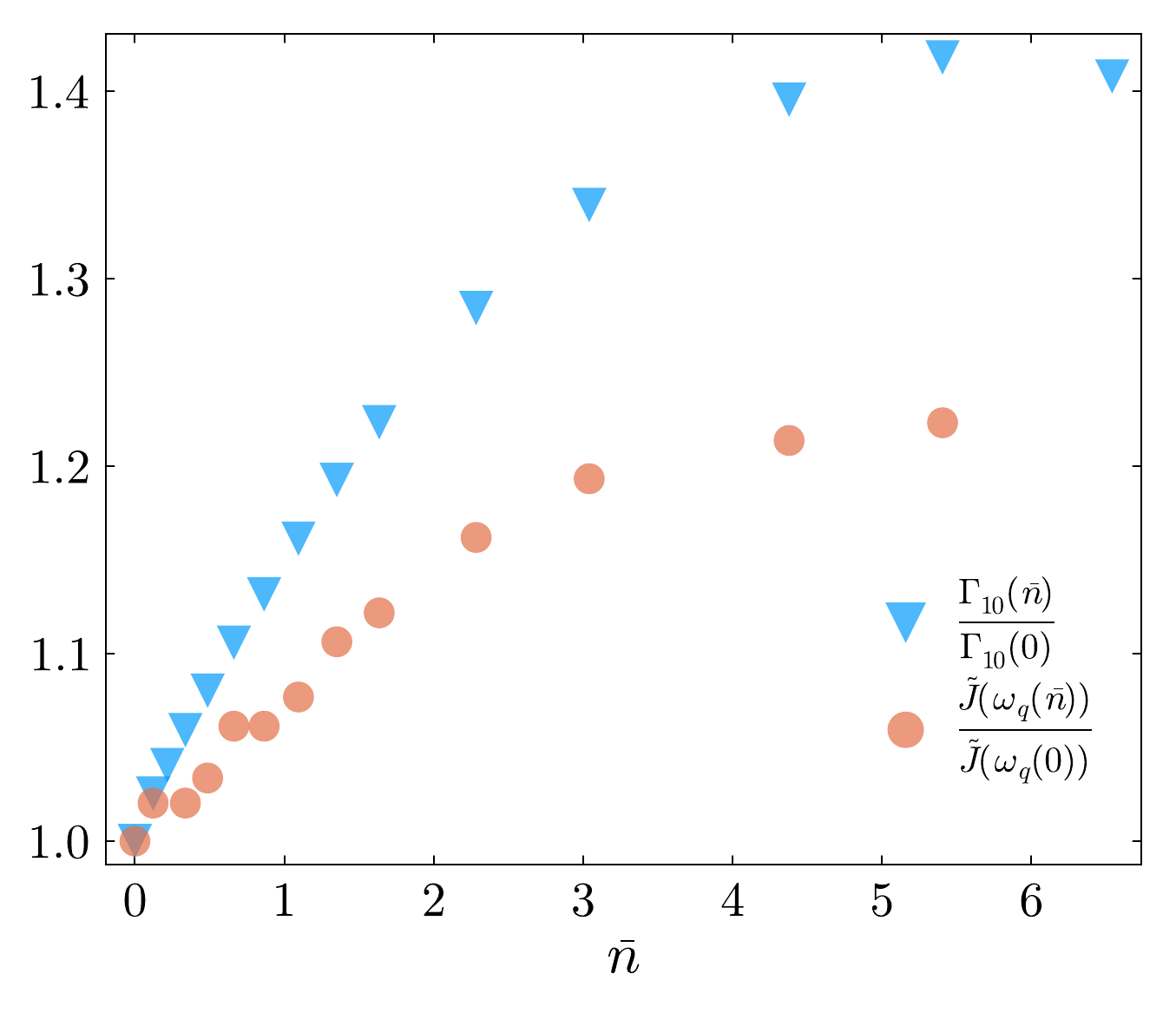}
    \caption{Comparison between $\Gamma_{10}(\bar n)/\Gamma_{10}(0)$ for the simulations of the Purcell-filtered spectral density $J_{\rm PF}(\omega)$ and the crude estimate obtained by taking $\Gamma_{10}(\bar n)$ proportional to $\frac{\tilde J (\omega_q(\bar n))}{\tilde J (\omega_q(0))}$, in terms of the effective spectral density $\tilde J_{\rm PF}(\omega)$ filtered by the resonator, as seen by the qubit.}
    \label{fig:effective}
\end{figure}

\paragraph{Estimate of the rates with the effective spectral density.--} To check for consistency, we show in \cref{fig:effective} a first-order estimate of the qubit decay rate using the effective spectral density function $\tilde J(\omega)$, which represents the bath filtered by the resonator. We compare the two ratios
\begin{align}\label{app:eq:forfig9}
    \frac{\Gamma_{10} (\bar n)}{\Gamma_{10} (0)}, \quad  \frac{\tilde J (\omega_q(\bar n))}{\tilde J (\omega_q(0))}, 
\end{align}
where $\omega_q(\bar n)$ is the ac Stark shifted frequency of the qubit, extracted from the simulation results of \cref{fig:spect_sigz}. This estimate accounts only for the change in the effective spectral density function at the shifted frequency. It neglects the renormalization of the transition matrix element of $\hat{\sigma}_x$, dressed by both the qubit-resonator interaction and by the drive. It also neglects the effect of qubit peak broadening with the readout drive photon number $\bar{n}$.

\section{C. Additional results for driven spectroscopy}

\begin{table}[]
    \centering
    \renewcommand{\arraystretch}{1.7} 
    \begin{tabular}{c|c|c|c}

         & $\, \, \omega_{a0}/(2\pi) \, \, $  & $\, \, \omega_{a1}/(2\pi) \, \, $ & $\, \, \bar \omega/(2\pi) \, \, $ \\
         
         \hline
       
       $\, \, J_{\rm Ohm}(\omega), \, \Delta>0\, \, $  &   $\, \, 7.546 \, \, $  &  $\, \, 7.47 \, \, $  & $\, \, 7.51 \, \, $ \\
       $\, \, J_{\rm Ohm}(\omega), \, \Delta<0\, \, $  &   $\, \, 5.26\, \, $  &  $\, \, 5.36 \, \, $  & $\, \, 5.31\, \, $ \\

       $\, \, J_{\rm flat}(\omega), \, \Delta>0\, \, $  &  $\, \, 7.55\, \, $   &  $\, \, 7.48 \, \, $  & $\, \, 7.52\, \, $ \\
       $\, \, J_{\rm PF}(\omega), \, \Delta>0\, \, $  &   $\, \, 7.546\, \, $  &  $\, \, 7.51\, \, $  & $\, \, 7.47\, \, $ \\
    \end{tabular}
    \caption{Frequencies in GHz extracted from the numerical calibration procedure for all the configurations considered.}
    \label{tab:calib}
\end{table}

\paragraph{Calibration.--}  The calibration procedure described in \cref{fig:calibration} was performed to select the drive frequencies for all the configurations considered. We report in \cref{tab:calib} the frequencies obtained.\\ 

\begin{figure}[t!]
    \centering
    \includegraphics[width=0.5\linewidth]{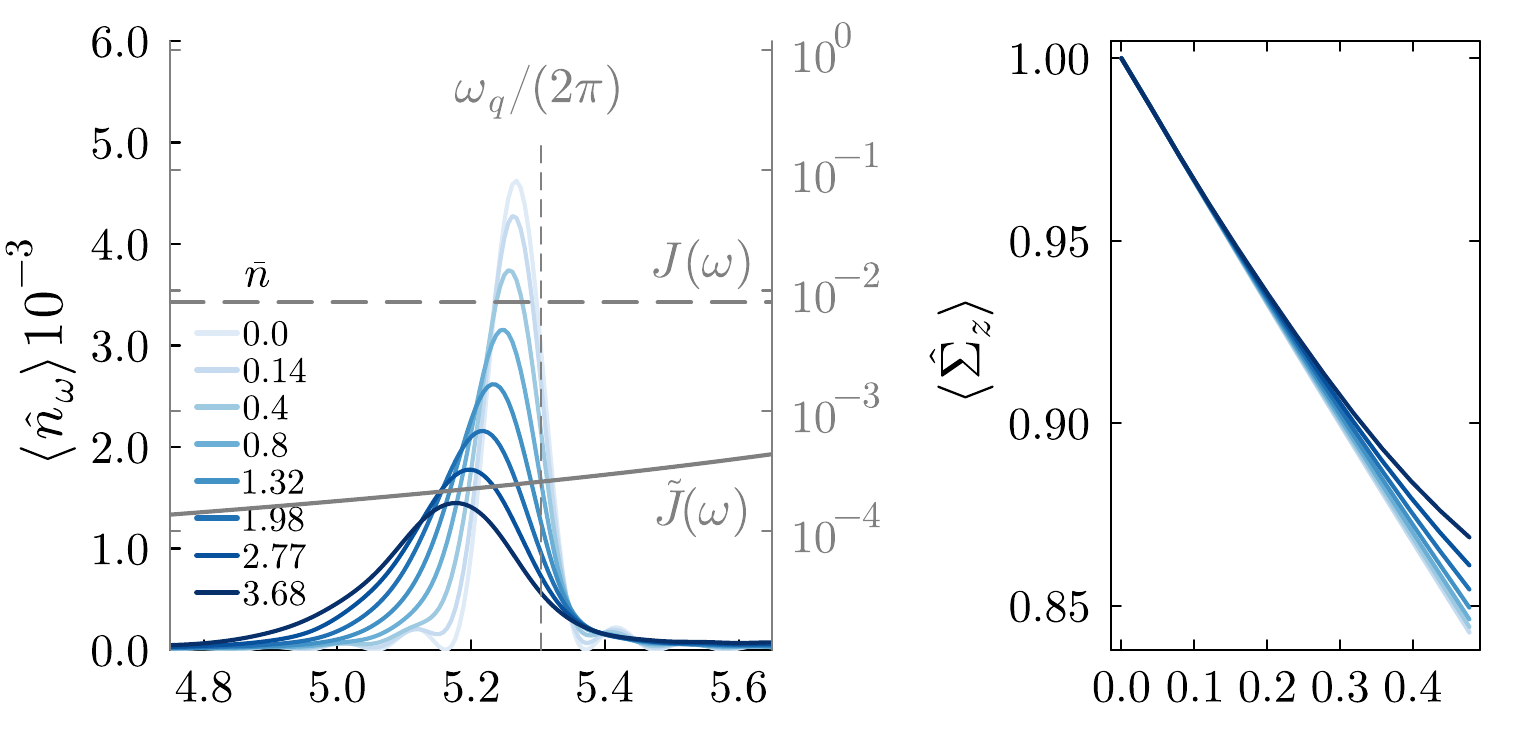} 
    \includegraphics[width=0.5\linewidth]{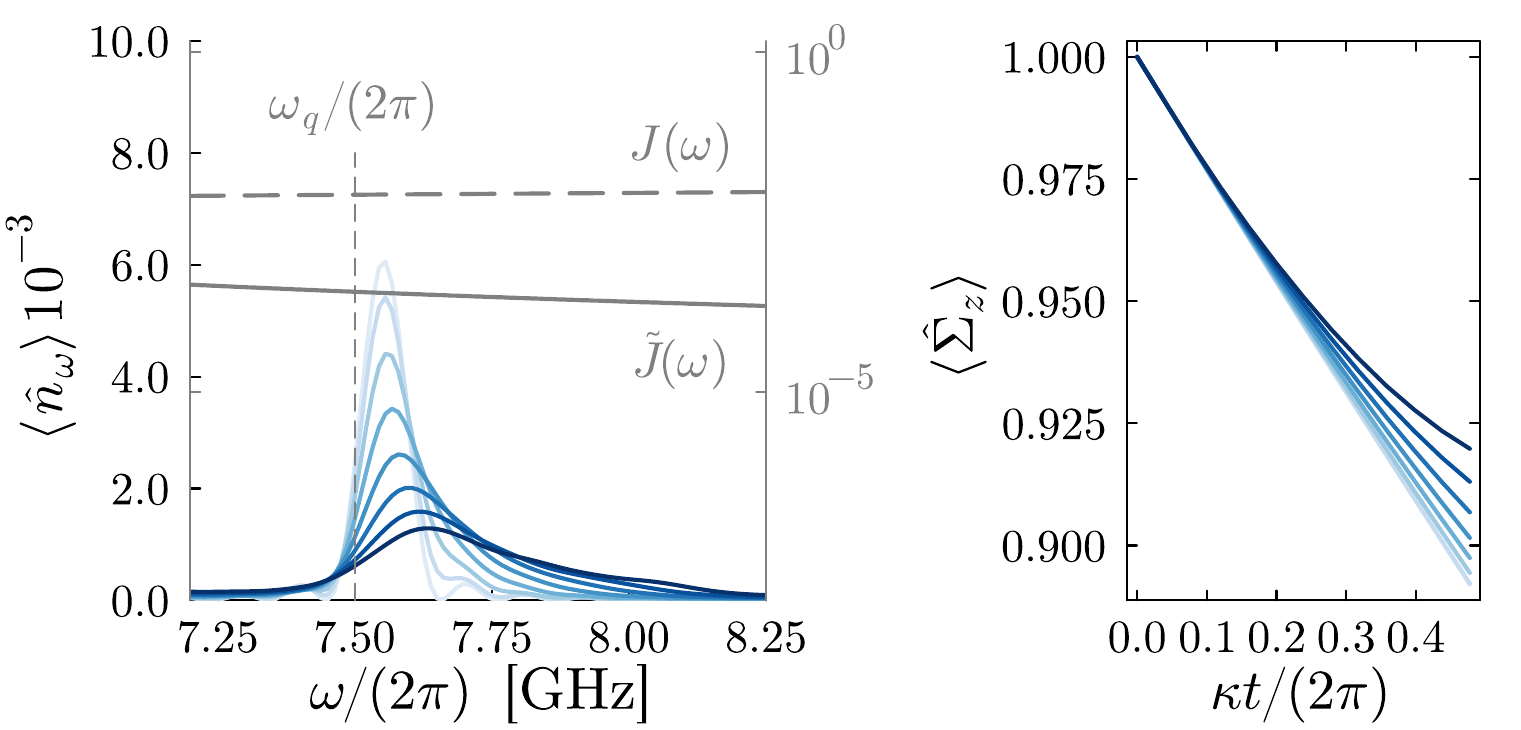}
    \caption{Same as \cref{fig:spect_sigz}, 
    for a flat $J(\omega) = J_{\rm flat}(\omega)$ (top) and Ohmic $J(\omega) = J_{\rm Ohm}(\omega)$ with $\Delta<0$ (bottom). The plots on the left show the bath mode occupations $\expval{\hat n_\omega}$ at final time $\kappa t/2\pi=0.5$, akin to a spectroscopy in the vicinity of the qubit frequency. The right plots show the time evolution of $\langle{\hat \Sigma_z}\rangle$, from which the relaxation rates of \cref{fig:t1_vs_nbar} have been extracted.}
    \label{fig:spect_appendix}
\end{figure}

\paragraph{Driven spectroscopy.--} We report in \cref{fig:spect_appendix} the results for the driven spectroscopy for both Ohmic spectral density $J_{\rm Ohm}(\omega)$ with $\Delta<0$, and the flat spectral density $J_{\rm flat}(\omega)$. In the Ohmic case with $\Delta<0$, the qubit is ac Stark shifted towards higher frequencies, according to the dispersive Hamiltonian prediction \cite{Blais_cQED}. The effective spectral density $\tilde{J}(\omega)$ gives an intuition on why $\Gamma_{10}$ decreases as $\bar n$ increases even though $J_{\rm Ohm}(\omega)$ grows with $\omega$: the upward Stark shift ($\Delta<0$) moves the effective qubit transition further away from the resonator frequency $\omega_a$, and the Lorentzian frequency filter induced by the resonator dominates the slope of $\tilde{J}(\omega)$, such that the net effect is the same as with $\Delta>0$.

\end{document}